\documentclass[aps,prb,twocolumn,floatfix,a4paper,amssymb,superscriptaddress,groupedaddress,longbibliography]{revtex4-2}   
\usepackage[english]{babel}
\usepackage[T1]{fontenc}
\usepackage{inputenc}
\usepackage{amsmath}
\usepackage{amsfonts}
\usepackage{amssymb}
\usepackage{graphicx}
\usepackage{dsfont}
\usepackage{tikz}
\usepackage[percent]{overpic}
\usepackage{xcolor}
\usepackage{physics}
\usepackage{float} 
\usepackage[caption = false]{subfig}
\usepackage{wrapfig}
\usepackage{pgfplots}
\usepackage{babelbib}
\usepackage{hyperref}
\usepackage{lipsum}
\usepackage{mathrsfs}
\usepackage{transparent}
\usepackage{amsthm}
\usepackage{listings}
\usepackage{eso-pic}
\usepackage{pgfplots}
\usepackage{dcolumn}   
\usepackage{bm}  
\usepackage{natbib}
\usepackage{xr}

\AtBeginDocument{%

}

\begin{document}
	\title{Hall Transport in Organic Semiconductors}	
	\author{Michel Panhans and Frank Ortmann}
	\affiliation{TUM School of Natural Sciences, Department of Chemistry, Technische Universit\"at M\"unchen, 85748 Garching b. M\"unchen, Germany}			

\begin{abstract}
We establish a universal theory to understand quasiparticle Hall effects and transverse charge-carrier transport in organic semiconductors. The simulations are applied to organic crystals inspired by rubrene and cover multiple transport regimes. This includes calculations of the intrinsic Hall conductivity in pristine crystals, which are connected with a simple description of semi-classical electron transport that involves the concept of closed electronic orbits in the band structure, which can be easily calculated in density functional theory.  
Furthermore, this framework is employed to simulate temperature-dependent longitudinal and transverse mobilities in rubrene. These simulations are compared to experimental findings, providing insights into these results by characterizing the non-ideality of the Hall effect due to the influence of vibrational disorder. We finally investigate the conditions for the observation of Shubnikov-de Haas oscillations in the longitudinal resistivity and quantized Hall plateaus in the transverse resistivity. A clear picture why this is not observed in rubrene is developed.
 These insights into classical and quantum Hall effects and their intermediates in organic semiconductors establish a blueprint for future explorations in similar systems.	
\end{abstract}
	\pacs{}
	\maketitle

\section{Introduction}
Investigating electronic transport by Hall measurements is of great interest for semiconductors and metals because Hall transport provides deep insights into the quasiparticle states, electron-transport mechanisms, and topological properties of band structures.\cite{vonKlitzing2020} The universality and efficiency of a theoretical description of Hall transport is indicated by the seamless transition between quantum and classical regimes and its applicability to challenging systems, respectively. High-mobility organic semiconductors (OSCs) are such systems, for which the classical Hall effect has been reported in rubrene single-crystal organic field-effect transistors (OFETs).~\cite{PodHall2005} Subsequent Hall measurements in rubrene \cite{Pod2013Hall,Frisbie2013,Frisbie2014Hall,Pod2016Hall,Pod20162Hall,Choi2018,Choi2020,Lee2014a} at moderately low temperatures and several other systems\cite{Choi2020,Sirringhaus2011,UemuraPRB2012,MinderAdvMat2012,YamagishiPRB2010,Fratini2020}, have been performed to study the nature of the charge carriers.
A two-dimensional hole gas in C$_8$-DNBDT-NW was measured down to temperatures around 10 K \cite{Kasuya2021}, in which the Hall mobilities indicate the existence of band transport. However, so far, no signature of a possible quantized Hall effect, like the integer quantum Hall effect (IQHE) \cite{Klitzing1986,Zhang2005a,Novoselov2007Science}, has been observed. The interest in the topological nature of these quantization phenomena, for which cryogenic temperatures lead to a step profile in the Hall resistance, opened a wide research field \cite{Klitzing1986,Novoselov2007Science,Thouless1982PRL,Haldane1988,KaneMele} but not yet in OSCs, where even numerical schemes for a quantitative description of Hall effects are only rarely reported in the literature \cite{Ishii2014,YuPRL2015}. Considering the breadth of experimental and theoretical developments of the Hall effect in various scientific contexts, its universal understanding covering all different facets including semi-classical band theories, Berry-phase physics \cite{Klitzing1986, Berry1984}, Boltzmann-transport approaches \cite{Ponce2021}, transient theories involving carrier localization \cite{Fratini2016,Fratini2017,Hutsch2022}, and carrier hopping \cite{Pollak1981,StepinaPRB2018,OberhoferChemRev2017}, etc. has not been established and seems out of reach. 

In this work, we develop such a universal theory and study the dc-Hall effect using an efficient time-domain representation \cite{Panhans2021PRL} for the electrical conductivity based on the general Kubo formula \cite{Panhans2021PRL,Kubo1957}. Within this quantum mechanical framework, we explicitly consider vibrational disorder arising from the electron-phonon-coupling (EPC) to low-energy molecular vibrations. We study different Hall-transport-regimes in rubrene by varying the strength of the disorder. While disorder can indeed vary in different experimental setups, our main purpose here is to generate more insights by this variation. First, we characterize the Hall conductivity for the states derived from the HOMO (highest occupied molecular orbital) of rubrene at vanishing disorder. 
By including vibrational disorder, we additionally investigate the temperature dependence of the Hall effect in bulk rubrene crystals and compare our results to experimental data.
It is shown that in presence of vibrational disorder, different transport states must contribute to longitudinal and transverse charge transport in full agreement with experimental findings. 
In the presence of weak disorder, in general the Hall conductivity can become quantized due to the formation of Landau bands \cite{Landau1930}, which shows the topological nature of the Hall effect owing to the intrinsic Berry curvature. \cite{Thouless1982PRL,Berry1984,NiuPRB1985,Kohmoto1985}  We determine the conditions under which Hall plateaus in the transverse resistivity and Shubnikov-de Haas (SdH) oscillations in the channel resistivity can be observed and when they break down.

\section{Theory}\label{sec:Theory}
\subsection{\uppercase{Hall-Transport Theory}}
We first introduce the essential ingredients of the developed Hall-transport theory and the numerical methods employed, which are necessary for both the conceptual understanding of Hall effects and for the practical evaluation of Hall responses in OSCs.
The theoretical basis is a time-domain approach to study linear responses based on the Kubo-formalism \cite{Panhans2021PRL}.  Within this framework, the electrical dc-conductivity tensor $\bm{\sigma}$ of effective non-interacting quasiparticles, reads
\begin{align}
	\begin{aligned}
	\sigma_{\alpha\beta}&=\frac{e^2}{2V}\int\limits_{0}^\infty dt \,e^{-\frac{t}{\tau}}\int\limits_{-\infty}^{\infty}dE\left[-\frac{1}{2}\frac{d f(E)}{d E}\frac{d^2}{dt^2}\mathscr{D}^+_{x_\alpha x_\beta}(E,t)\right.\\&\left.+\frac{1}{\hbar}f(E)\frac{d}{dt}\mathscr{D}^-_{x_\alpha x_\beta}(E,t)\right],\label{DC-conductivity}
	\end{aligned}
\end{align}
with the Fermi function $f(E) = (e^{\beta(E-\mu)}+ 1)^{-1}$, the (phenomenological) relaxation time $\tau$, and the correlation functions
\begin{align}
\mathscr{D}^+_{x_\alpha x_\beta}(E,t)&=\text{Tr}(\delta(E-\hat{H})\left\{\Delta \hat{x}_\alpha (t),\Delta \hat{x}_\beta (t)\right\}),\label{DAF}\\					
\mathscr{D}^-_{x_\alpha x_\beta}(E,t)&=-i\text{Tr}(\delta(E-\hat{H})\left[\Delta \hat{x}_\alpha (t),\Delta \hat{x}_\beta(t)\right]),\label{DCF} 		
\end{align}				
in which the square brackets and the curly brackets denote the commutator and the anticommutator, respectively. The time evolution of $\Delta \hat{x}_\alpha (t) = \hat{x}_\alpha (t)-\hat{x}_\alpha (0)$ for the Cartesian direction $\alpha$ is driven by the Hamiltonian $\hat{H}$, describing effectively non-interacting quasiparticles (see Appendix \ref{app:A} and \ref{app:B} for the detailed derivation). 
The symmetric and the antisymmetric parts of the conductivity tensor components $\sigma_{\alpha\beta}^{\text{s(as)}}=(\sigma_{\alpha\beta}+(-)\sigma_{\beta\alpha})/2$ then read
\begin{align}
\sigma_{\alpha\beta}^{\text{s}}&=-\frac{e^2}{4V}\lim\limits_{\tau\to\infty}\int\limits_{0}^\infty dt \,e^{-\frac{t}{\tau}}\int\limits_{-\infty}^{\infty}dE\frac{df(E)}{d E}\frac{d^2}{dt^2}\mathscr{D}^+_{x_\alpha x_\beta}(E,t),\label{symmetric part dc-conductivity}\\
\sigma_{\alpha\beta}^{\text{as}}&=\frac{e^2}{2V\hbar}\lim\limits_{\tau\to\infty}\int\limits_{0}^\infty dt \,e^{-\frac{t}{\tau}}\int\limits_{-\infty}^{\infty}dEf(E)\frac{d}{dt}\mathscr{D}^-_{x_\alpha x_\beta}(E,t).\label{anti-symmetric part dc-conductivity}
\end{align}
The Hall conductivity is given by the antisymmetric part of the tensor, i.e., $\sigma_\text{H}=\sigma_{\alpha\beta}^\text{as}$. Due to the general Onsager-Casimir relations \cite{Onsager1931,CasimirRMP1945,Luo2020} for the electrical conductivity $\sigma_{\alpha\beta}(B_\gamma)=\sigma_{\beta\alpha}(-B_\gamma)$, only the antisymmetric part $\sigma_{\alpha\beta}^\text{as}$ describes a field-induced Hall voltage $V_\text{H}=[V_\text{H} (B_\gamma )-V_\text{H} (-B_\gamma)]/2$, while the symmetric part $\sigma_{\alpha\beta}^\text{s}$ is invariant under reversal of the magnetic field.
We numerically evaluate Eq.~\eqref{DC-conductivity}, using a linear-scaling quantum-transport approach \cite{Fan2021a}, the Lanczos algorithm, and a Chebyshev expansion for the determination of the functions $\mathscr{D}^\pm_{x_\alpha x_\beta}(E,t)$. 
From the computed dc-conductivity tensor $\boldsymbol{\sigma}$, we obtain the resistivity tensor by inversion, $\boldsymbol{\rho}=\boldsymbol{\sigma}^{-1}$.\\
\subsection{\uppercase{Relation to other Hall-transport theories}}
In this theoretical section, we shortly discuss how the present work includes other known Hall-transport approaches used in the past such as the description of the Hall conductivity in terms of the Berry curvature.\cite{Thouless1982PRL}
From the present approach starting from Eq.~\eqref{DC-conductivity} (see Appendix \ref{app:D} for details), we obtain the Hall conductivity as
\begin{align}
\begin{aligned}
\sigma_\text{H}
&=\frac{ie^2\hbar}{V}\lim\limits_{\tau\to\infty}\sum\limits_{mn}f(\varepsilon_n)\frac{\left(v^\alpha_{nm} v^\beta_{mn}-v^\beta_{nm}v^\alpha_{mn}\right)}{(\varepsilon_n-\varepsilon_m)^2+\left(\frac{\hbar}{\tau}\right)^2},
\end{aligned}\label{Berry}
\end{align}
with $v^{\alpha}_{nm}$ being the matrix elements of the velocity operator expanded in the quasiparticle eigenbasis of a proper quasiparticle Hamiltonian with eigenenergies $\varepsilon_n$. In the limit $\tau\to\infty$, the Hall conductivity is obtained from the thermally averaged Berry curvature, which is the key property in the theory of the IQHE.

For the symmetric part of the conductivity $\sigma^\text{s}_{\alpha\beta}$, we obtain from Eq.~\eqref{DC-conductivity} (see Appendices \ref{app:B} and \ref{app:D}) in the limit $\tau\to\infty$:
 \begin{align}
 \begin{aligned}
 \sigma_{\alpha\beta}^{\text{s}}
 &=\frac{\beta e^2\pi\hbar}{2V}\sum\limits_{mn}f(\varepsilon_n)\left(1-f(\varepsilon_n)\right)\delta(\varepsilon_n-\varepsilon_m)\\&\quad\times\left(v^\alpha_{nm} v^\beta_{mn}+ v^\beta_{nm}v^\alpha_{mn}\right).
 \end{aligned}
 \end{align}
 For $\alpha=\beta$, this symmetric part becomes equivalent to the Kubo-Greenwood-type formula for longitudinal charge transport \cite{Greenwood_1958,Fan2021a}.

In addition, we prove in Appendix \ref{app:E} that Eq.~\eqref{DC-conductivity} is also equivalent to the Kubo-Bastin\cite{BASTIN19711811} formula and to the Kubo-Bastin-Str\v{e}da formula \cite{Streda_1982}, which establishes additional exact relationships to other numerical schemes used to study charge transport in (topological) condensed matter systems.\cite{Garcia2015PRL,Mitscherling2020,CastroPRL2024}. 
More specifically, the Kubo-Bastin formula \cite{BASTIN19711811} describes the dc-conductivity in the form:
\begin{align}
\sigma_{\alpha\beta}&=\frac{e^2}{V}\int\limits_{-\infty}^\infty dE f(E)C_{\alpha\beta}(E),\label{Kubo-Bastin}
\end{align}
with the correlation function
\begin{align}
C_{\alpha\beta}(E)&=i\hbar\text{Tr}\left[\hat{v}_\alpha\delta(E-\hat{H})\hat{v}_\beta\frac{d\hat{G}^+(E)}{dE}-h.c.\right],
\end{align}
which depends on the one-particle Green's function operator $\hat{G}^\pm(E)=\lim_{\eta\to 0}(E-\hat{H}\pm i\eta)^{-1}$.
The equivalence of Eq.~\eqref{DC-conductivity} with the Kubo-Bastin and the Kubo-Bastin-Str\v{e}da formula is obtained from the decomposition of the correlation function $C_{\alpha\beta}(E)$ into symmetric and antisymmetric parts $C_{\alpha\beta}^\text{s(as)}(E)=(C_{\alpha\beta}(E)+(-)C_{\beta\alpha}(E))/2$ yielding
\begin{align}
&	\begin{aligned}
C_{\alpha\beta}^\text{s}(E)&=\pi\hbar\frac{d}{dE}\text{Tr}\left[\delta(E-\hat{H})\hat{v}_\alpha\delta(E-\hat{H})\hat{v}_\beta\right],
\end{aligned}
\\
&\begin{aligned}
C_{\alpha\beta}^\text{as}(E)&=i\hbar\text{Tr}\left[\delta(E-\hat{H})\left(\hat{v}_\alpha\frac{1}{(E-\hat{H})^2}\hat{v}_\beta-h.c.\right)\right].
\end{aligned}
\end{align}
From this decomposition, we obtain the same result for $\sigma_{\alpha\beta}^\text{s}$ and $\sigma_\text{H}$ as derived from Eq.~\eqref{DC-conductivity} and thus demonstrate the equivalence to the Kubo-Bastin formula in Eq.~\eqref{Kubo-Bastin} (see Appendices \ref{app:D} and \ref{app:E} for details).
To compare Eq.~\eqref{DC-conductivity} with the Kubo-Bastin-Str\v{e}da formula, the Hall conductivity is decomposed into two terms
$\sigma_\text{H}=\sigma_\text{H}^\text{I}+\sigma_\text{H}^\text{II}$,
where $\sigma_\text{H}^{\text{II}}$ is given by
\begin{align}
\sigma_\text{H}^\text{II}=\frac{e^2}{2V}\int\limits_{-\infty}^\infty dE\,f(E)\text{Tr}\left[\frac{d}{dE}\delta(E-\hat{H})\left(\hat{x}_\alpha\hat{v}_\beta-\hat{x}_\beta\hat{v}_\alpha\right)\right],
\end{align}
 which explicitly appears in the Str\v{e}da formula and has previously been used to explain the quantization of the Hall conductivity.\cite{Streda_1982} This contribution to the Hall conductivity is also relevant to explain the intrinsic Hall conductivity in the disorder-free case (see below) and thus is, interestingly, at the origin of the semi-classical version of the Hall conductivity discussed in Sect.~\ref{subsec:intrinsic}. 
 
 In addition to covering previous theories, Eq.~\eqref{DC-conductivity} provides us with the full quantum-dynamical time evolution of the involved transport correlation functions. This permits investigating any transport regime, which otherwise demands conceptually different theories used previously. It also offers detailed information on, e.g., the formation of topological states on the time scale of localization, and scattering effects.

\subsection{\uppercase{Effective quasiparticle Hamiltonian and simulation geometry for rubrene}}
We study the Hall effect for a three-dimensional rubrene crystal with lateral dimensions of 0.06$\mu$m$\times$0.8$\mu$m$\times$1.6$\mu$m, including around 10$^\text{8}$ coupled hole orbitals. 
We focus on the HOMO-derived states as an example, as holes are the usual carrier type in rubrene and similar OSCs.
In this setup, the $\alpha$-direction (see Fig.~\ref{fig:fig1}(a)) is aligned with the high-mobility direction of the rubrene crystal, while the $\beta$-direction is perpendicular to $\alpha$ and lies also inside the herringbone plane. 
The lattice parameters of the crystal are given by $a=$ 26.789\,\AA \, ($\gamma$-direction), $b=$ 7.170\,\AA \, ($\alpha$-direction see Fig.~\ref{fig:fig1}), and $c=$ 14.211\,\AA \, ($\beta$-direction) of an orthorhombic Bravais lattice and are taken from the literature \cite{Ordejon2017}. 
The four rubrene molecules arrange in a herringbone structure in the $\alpha\beta$-plane with the centre-of-mass coordinates $\text{A}=\left(0,0,0\right)$,  $\text{B}=\left(\frac{a}{2},0,\frac{c}{2}\right)$,  $\text{C}=\left(0,\frac{b}{2},\frac{c}{2}\right)$ and $\text{D}=\left(\frac{a}{2},\frac{b}{2},0\right)$. 
A suitable tight-binding expression for the four bare electronic bands with a minimal number of electronic transfer integrals reads  
\begin{align}
	\begin{aligned}
		\varepsilon(\mathbf{k},\pm,\pm)&=2\varepsilon_{\text{AA}+\mathbf{b}}\cos(k_\alpha b)+2\varepsilon_{\text{AA}+2\mathbf{b}}\cos(2k_\alpha b)\\&\quad\pm 4\varepsilon_{\text{AC}}\cos\left(\frac{k_\alpha b}{2}\right)\cos\left(\frac{k_\beta c}{2}\right)\\&\quad\pm 4\varepsilon_{\text{AB}}\cos\left(\frac{k_\gamma a}{2}\right)\cos\left(\frac{k_\alpha b}{2}\right),
	\end{aligned}\label{TB-band structure rubrene}
\end{align}       
with the transfer integrals $\varepsilon_{ij}$ taken from prior work. \cite{Ordejon2017}. 

To model hole transport in rubrene crystals, these parameters enter a Hamiltonian of the form 
\begin{align}
\hat{H}=\sum\limits_{\left<i,j\right>}e^{-i\varphi_{ij}}\tilde{P}\left(\varepsilon_{ij}+V_{ij}(T)\right)\hat{c}_i^\dagger\hat{c}_j+\sum\limits_{i}V_{ii}(T)\hat{c}_i^\dagger\hat{c}_i, \label{effective polaron Hamiltonian}
\end{align}
which is derived from a Holstein-Peierls Hamiltonian. 
Eq. \eqref{effective polaron Hamiltonian} includes, besides the above $\varepsilon_{ij}$ between the molecular orbitals, an effective description of the EPC, where the matrix elements $V_{ii}(T)$ and $V_{ij}(T)$ (see below) describe a temperature-dependent vibrational disorder potential caused by the coupling to low-energy vibrational modes. It also includes a polaron renormalization factor $\tilde{P}$ caused by the coupling to high-energy vibrational modes
(see Refs.~\cite{Hutsch2022,Hannewald2004,Ortmann.2009b,MerkelPRB2022} for details). The Peierls phase $\varphi_{ij}$ \cite{Luttinger1951PRL} describes the interaction of the holes with a homogeneous magnetic field
and is generically defined as \cite{Luttinger1951PRL}
\begin{align}
	\varphi_{ij}=\frac{e}{\hbar}\int_{\mathbf{r}_i}^{\mathbf{r}_j}d\mathbf{r}\cdot\mathbf{A}(\mathbf{r}),
\end{align}
with the vector potential $\mathbf{A}(\mathbf{r})=Bx_\beta\mathbf{e}_\alpha+B'x_\alpha\mathbf{e}_\beta$ and the resulting magnetic field $\mathbf{B}=B_\gamma\mathbf{e}_\gamma$ with the constant amplitude $B_\gamma=B-B'$.
The contributions to the model Hamiltonian that arise from EPC are given by (see e.g., Ref.~\cite{MerkelPRB2022})
\begin{align}
V_{ij}(T)&=\sum\limits_\lambda\hbar\omega_\lambda g^\lambda_{ij}\sqrt{1+2n_\lambda}\left(\frac{\phi_i^\lambda+\phi_j^\lambda}{2}\right),\\
\tilde{P}&=e^{-\sum\limits_\xi \left(g_{ii}^\xi\right)^2\left(1+2n_\xi\right)}. \label{vibrational disorder potential}
\end{align}
Here, $\phi_i$ are Gaussian random numbers describing the vibrational disorder of low-energy modes, which arises from tracing out these degrees of freedom. $\hbar\omega_{\lambda}$ and $\hbar\omega_{\xi}$ are the vibrational mode energies of the low-energy and high-energy modes, respectively. $n_\lambda=1/(e^{\beta\hbar\omega_\lambda}-1)$ and $n_{\xi}$ (defined accordingly) are their thermal occupations, while $g_{ij}^{\lambda}$ and $g_{ij}^{\xi}$ are the respective coupling constants to the electronic states.
The vibrational disorder is based on \textit{ab initio} material parameters taken from Ref.~\cite{Ordejon2017} and is a static Gaussian disorder, which modifies the onsite energies $\varepsilon_{ii}$ and the transfer integrals $\varepsilon_{ij}\in\left\{\varepsilon_{\text{AA}+\mathbf{b}},\varepsilon_{\text{AA}+2\mathbf{b}},\varepsilon_{\text{AB}},\varepsilon_{\text{AC}}\right\}$ of the rubrene crystal.  The mean strength of the vibrational disorder $V_{ij}(T)$ is determined by its standard deviation averaged over all HOMO states, which is obtained from the EPC material parameters as 
\begin{align}
\begin{aligned}
V^2_{\text{EPC}}(T)
&=\sum\limits_{\lambda}\left(\hbar\omega_{\lambda}\right)^2\left(1+2n_{\lambda}\right)\left[(g_{ii}^{\lambda})^2+\frac{1}{2}\sum\limits_{j\neq i}g^{\lambda}_{ij}g^{\lambda}_{ji}\right]\\
&=V^2_{0}(T)+V^2_{\text{AA}+\mathbf{b}}(T)+V^2_{\text{AA}+2\mathbf{b}}(T)\\&\quad+2V^2_{\text{AC}}(T)+2V^2_{\text{AB}}(T), \label{thermal disorder strength}
\end{aligned}
\end{align}
where the sum over $j$ only runs over the next neighbors of $i$ with respect to the corresponding transfer integral. The sum over $\lambda$ only includes low-frequency modes below a mode energy of 75 meV. 
The temperature dependence of the vibrational disorder potential is co-determined by the thermal occupation $n_\lambda$ of mode $\lambda$. 

In Tab.~\ref{tab:material parameters} (Appendix \ref{app:F}), we summarize the energy parameters used in the effective polaron model in Eq.~\eqref{effective polaron Hamiltonian}.
We note that the value for $V_{\text{AA}+2\mathbf{b}}(T)$ is set to zero since the EPC constants $g_{\text{AA}+2\mathbf{b}}^{\lambda}$ have not been considered in Ref.~\cite{Ordejon2017} for the second-neighbor transfer integral $\varepsilon_{\text{AA}+2\mathbf{b}}$. From Tab.~\ref{tab:material parameters}, we see that the vibrational disorder is dominated by the intramolecular contribution $V_0(T)$ and has a total disorder strength of $V_\text{EPC}^\text{300K}=57.5$ meV.
The polaron renormalization $\tilde{P}$ obtained from the remaining high-frequency modes is evaluated to $\tilde{P}=0.72$, which reduces the transfer integrals and the intermolecular vibrational disorder.

Finally, the cyclotron energy close to the HOMO band edge reads $\hbar\omega_\text{cyc}=\hbar eB_\gamma/m_\text{cyc}$ with the inverse cyclotron mass
\begin{align}
\begin{aligned}
m_\text{cyc}^{-1}&=\frac{\tilde{P} bc}{\hbar^2}\sqrt{\varepsilon_{\text{AC}}\left(2\varepsilon_{\text{AA}+\mathbf{b}}+8\varepsilon_{\text{AA}+2\mathbf{b}}+\varepsilon_{\text{AC}}+\varepsilon_{\text{AB}}\right)},
\end{aligned}
\end{align}
which is derived from the pristine (but renormalized) HOMO band structure in Eq.~\eqref{TB-band structure rubrene}. The cyclotron energy near the HOMO band edge amounts to 0.51 (5.1) meV for a magnetic field strength of 6 T (60 T).

The results derived from the proposed theoretical framework in Sect.~\ref{sec:Results} below, which encompasses investigations of three transport scenarios, illustrate distinct regimes of Hall transport. These include semi-classical, quantum, topological, and localization effects, all integrated within a single framework. There, to illustrate the seamless transition between these regimes, we vary the EPC-induced disorder strength from zero to its full material-parameter-based values.

\section{Results}\label{sec:Results}
\subsection{\uppercase{The intrinsic Hall conductivity}}\label{subsec:intrinsic}
\begin{figure*}[th]
	\centering
	\includegraphics[width=0.8\linewidth]{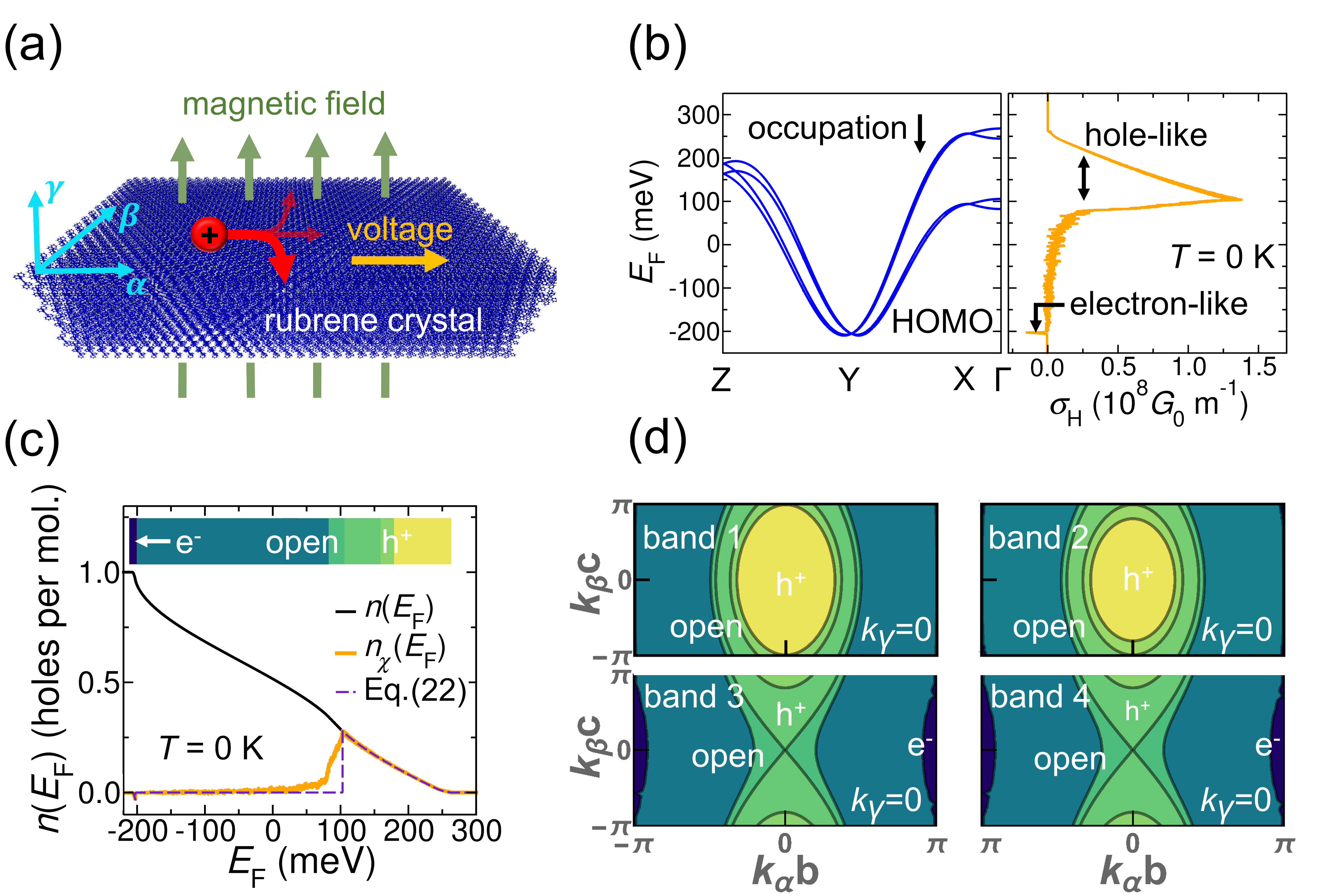}
	\caption{Hall effect in rubrene at vanishing disorder. (a) Hall setup for bulk crystals, where a homogeneous magnetic field is applied perpendicular to the herringbone plane ($\alpha\beta$-plane). (b) (left panel) Band structure of the four HOMO bands. (right panel) Energy-resolved Hall conductivity $\sigma_{\text{H}}$ at 0 K.  The axes are flipped ($\sigma_{\text{H}}\leftrightarrow E_\text{F}$) to compare both panels more easily. (c) Energy-resolved density $n_\chi$ at 0 K. The total charge-carrier concentration $n$ (black) and the analytical result from Eq.~\eqref{approximate sigma_H} (dashed indigo line) are plotted for comparison. The colored bar in the inset defines different regimes and their energy separation.  (d): Contour plot of the Fermi surface at $k_\gamma$=0 for all four bands at different energies. The contour lines are defined by energies of the colored bar in (c). Hole-like, electron-like, and open-orbit domains are indicated.}
	\label{fig:fig1}
\end{figure*}

The first Hall-transport scenario is the intrinsic limit of vanishing disorder, where we focus only on the Hall-transport properties stemming from the HOMO-band structure. In this first scenario, the coupling to low-energy vibrations, i.e., the vibrational disorder is neglected, while the coupling to high-energy vibrations is included via a polaron renormalization factor of $\tilde{P}=0.72$. 
We numerically evaluate the conductivity tensor in the herringbone plane ($\alpha\beta$-plane, see Fig.~\ref{fig:fig1}(a)) and calculate the Hall conductivity $\sigma_{\text{H}}$ at a magnetic field strength of $B_\gamma=6$ T, which is a typical modest value used in experiments.
In Fig.~\ref{fig:fig1}(b) (right panel), the energy-resolved Hall conductivity (at 0 K for simplicity) is compared to the pristine band structure of the four HOMO bands (left panel in Fig.~\ref{fig:fig1}(b)) in rubrene bulk crystals. The results show an intrinsic Hall effect in the absence of disorder, which we now analyze in more detail. 

We make two major observations: (i) The Hall conductivity is monotonically increasing from the top of the HOMO band around $E_\text{F}=$ 250 meV. However this trend persists only down to a Fermi energy of around 100 meV. In this energy region, it shows a clear hole-like behavior, where the magnetic-field response agrees with the usual right-hand rule (see Fig.~\ref{fig:fig1}(a)) as described by a fully classical model of the Hall effect. At lower energy ($E_\text{F}\leq100$ meV), $\sigma_{\text{H}}$ drops rapidly to zero or turns even negative, which cannot be explained by a reasonable classical model and requires a quantum mechanical description. (ii) Despite the absence of disorder in the present scenario, one cannot observe any quantization effect in $\sigma_\text{H}$ even at $T=0K$. 

To understand these observations and the distinct energy dependence of the Hall conductivity in Fig.~\ref{fig:fig1}(b), we analyze it by relating $\sigma_\text{H}$ to the total carrier concentration $n$ to quantify the number of hole states that contribute to the Hall effect. We thus define the density $n_\chi$ that is characteristic for $\sigma_\text{H}$ according to
\begin{align}
en_\chi=\sigma_\text{H} B_\gamma.	\label{en_chi}										
\end{align}
$n_\chi$ is compared to the total carrier concentration $n$ in Fig.~\ref{fig:fig1}(c). This shows that at low concentrations $n_\chi$ agrees with $n$ for $E_\text{F}\geq103$ meV (or $n\leq0.28$ holes per molecule). In addition, we find that for $n\leq 0.28$, $en_\chi$ is also equal to the Hall density $en_\text{H}$, which implies $n_\text{H}=n_\chi=n$ and thus describes an ideal Drude-like Hall effect $(n_\text{H}=n)$ of the holes with a Hall resistance of $\rho_\text{H}=\sigma_\text{H}^{-1}$. 
 At lower Fermi energy, $n_\chi$ decreases rapidly (between 80 meV $\leq E_\text{F}\leq100$ meV) and almost vanishes for even lower energies.
Since we do not observe any quantization effects in $\sigma_\text{H}$ and $\rho_\text{H}$, we conclude that the obtained energy dependence of the Hall conductivity is an intrinsic property of the pristine band structure and should coincide with a semi-classical electron-transport theory.~\cite{VonN.W.Ashcroft;N.D.Mermin1976}

Motivated by these numerical findings and going beyond the specific rubrene crystals, we have derived a  simple and general analytic expression for the intrinsic Hall conductivity from the full quantum description in Eq.~\eqref{DC-conductivity}. In absence of disorder and for a purely two-dimensional system (taken for simplicity), the intrinsic Hall conductivity yields the semi-classical expression
\begin{align}
	\sigma_{\text{H}}^\text{sc}=\frac{eN}{B_\gamma V}\int\limits_{-\infty}^\infty dE\,f(E)\frac{d}{dE}\frac{A_e(E)-A_
		h(E)}{\Omega_\text{BZ}}. \label{sigma_H sc}
\end{align}
This result involves the area difference $A_e(E)-A_h(E)$ enclosed by all semi-classical quasiparticle orbits (both electron- and hole-like) at energy $E$ in the Brillouin zone (BZ). $\sigma_{\text{H}}^\text{sc}$ furthermore depends on the  total number of electronic states $N$, the chemical potential $\mu$,  while $\Omega_\text{BZ}$ is the BZ volume (see Appendix \ref{app:C} for details of this derivation). 
For the particular case of rubrene, the Hall conductivity at zero temperature can further be approximated from Eq.~\eqref{sigma_H sc} as:
\begin{align}
	\sigma_{\text{H}}=\frac{e}{B_\gamma}\left[n\left(1-\Theta(E_\text{F}-E_h)\right)-n_e\Theta(E_\text{F}-E_e))\right], \label{approximate sigma_H}
\end{align}
with the Heaviside step function $\Theta(E)$.
The energies $E_h\approx 103$ meV and $E_e\approx -202$ meV are introduced as the critical energies that separate energy domains where the Fermi surface describes closed hole-like and electron-like orbits.
Fig.~\ref{fig:fig1}(d) illustrates these orbits as contours for $k_\gamma=0$. Eq. \eqref{approximate sigma_H} describes the fact that for a Fermi energy of $E_\text{F}>E_h$ an ideal hole-like Hall effect is observed, i.e.,  $\sigma_{\text{H}}=en/B_\gamma$ and for $E_\text{F}<E_e$ an ideal electron-like Hall effect, i.e., $\sigma_{\text{H}}=-en_e/B_\gamma$ (with the electron concentration $n_\text{e}=1-n$) is observed. For $E_e<E_\text{F}<E_h$ the Hall conductivity vanishes. This is because the orbits are open in full agreement with the numerical results of $en_\chi$ in Fig.~\ref{fig:fig1}(c). The closure of the orbits with changing energy can be seen in Fig.~\ref{fig:fig1}(d).

At this point we should emphasize the quality of these findings.
Within the full quantum-mechanical treatment of $\sigma_{\text{H}}$ (including the possible topological nature of the Hall conductivity) we have shown here that, in the absence of quantization phenomena and under conditions of sufficiently low disorder, the intrinsic Hall conductivity can be effectively explained. This explanation relies exclusively on analyzing the semi-classical orbits within the pristine band structure of the specified material (at zero magnetic field). 
This picture also captures transitions between ideal and non-ideal Hall effects.
Moreover, we find that the absence of quantization of the Hall conductivity is due to the relatively small energy spacing of the Landau bands at moderate fields of $B_\gamma=6$ T. The level spacing is competing with the band dispersion along the direction of the magnetic field, which is larger in the present case. This establishes the conditions under which such semi-classical responses may arise.
We finally note that Eq.~\eqref{sigma_H sc} can also be derived by assuming \emph{a priori} a semi-classical regime of the Hall effect. \cite{LifshitzJETP1956,AraiPRB2009} It is closely related to the presence of Lifshitz transitions \cite{lifshitz1960anomalies} in the pristine band structure describing changes in the topology of the Fermi surface when changing the Fermi energy. Thus, we find this regime to be a part of the present and more general Kubo approach.

\subsection{\uppercase{Temperature-dependent Hall effect in bulk rubrene crystals}}\label{subsec:temperature-Hall}
We continue our analysis of the Hall effect in rubrene by studying its temperature dependence at a magnetic-field strength of 6 Tesla. In addition to Sect.~\ref{subsec:intrinsic}, we now include the vibrational disorder based on our molecular parameters for the EPC in rubrene. 
To characterize the present Hall-transport regime, we calculate the Hall mobility $\mu_\text{H}$, the channel mobility $\mu$, and the Hall density $en_\text{H}$, which can be measured in Hall probes. More explicitly, these quantities are obtained from the resistivity components, yielding
\begin{align}
\mu_{\text{H}}=\frac{\rho_{\alpha\alpha}}{B_\gamma\rho_\text{H}},&& \mu=\frac{1}{en\rho_{\alpha\alpha}},&&en_\text{H}=\frac{B_\gamma}{\rho_\text{H}},
\end{align}
where $\rho_{\alpha\alpha}=\rho$ is the resistivity along the high-mobility direction and $\rho_{\text{H}}$ is the Hall resistance along the $\beta$-direction (see Fig.~\ref{fig:fig1}(a) for comparison). The resistivities $\rho_{\alpha\alpha}$ and $\rho_\text{H}$ are obtained from inversion of the full dc-conductivity tensor obtained from Eq.~\eqref{DC-conductivity}.

 \begin{figure}[t!]
	\centering
	\includegraphics[width=0.95\linewidth]{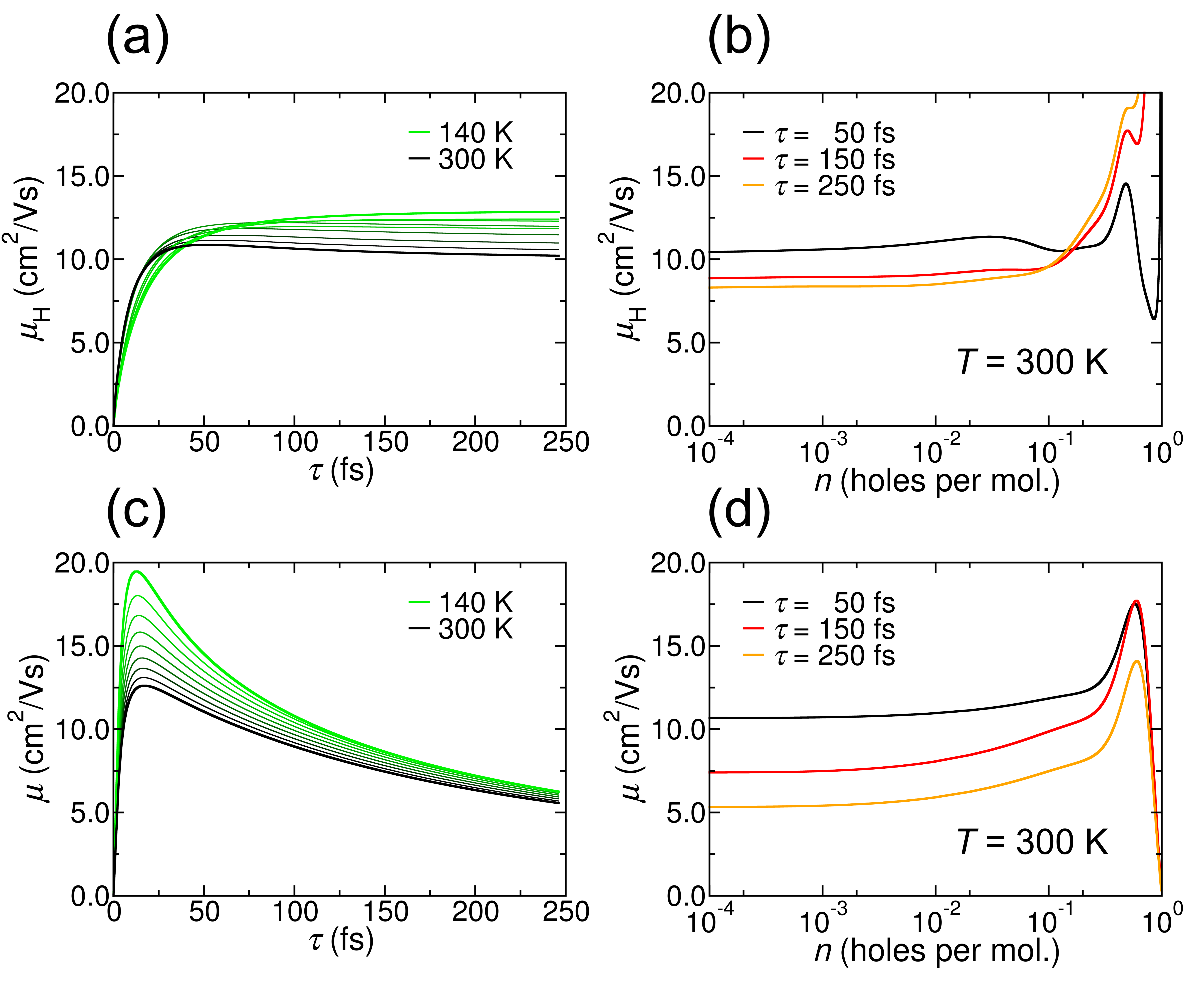}
	\caption{Temperature-dependent Hall transport in bulk rubrene crystals at 6 T. (a), (c) Time dependence of  the Hall mobility and the channel mobility for temperatures between 140 K and 300 K in steps of 20 K.
		(b), (d) Concentration dependence of the Hall mobility and channel mobility for fixed $\tau$ of 50 fs, 150 fs, and 250 fs at 300K. In (a) and (c), the carrier concentration is fixed to $n=2.5\times10^{-4}$ holes per molecule. }
	\label{fig:fig2}
\end{figure}

To get detailed insights into this Hall-transport regime, we analyze the dependencies of $\mu_\text{H}$ and $\mu$ on the relaxation time $\tau$ and on the carrier concentration $n$.
In Fig.~\ref{fig:fig2}(a), we plot the Hall mobility as a function of $\tau$, which turns out to be almost constant for $\tau\ge$ 50 fs and varies only slightly with temperature. On the other hand, the channel mobility $\mu$ in Fig.~\ref{fig:fig2}(c) has a stronger dependence on $\tau$ and decays at larger times.  Note that large $\tau$ can be associated to weaker external relaxation effects.
This decay with $\tau$ is a consequence of the vibrational disorder entering the Hamiltonian in Eq.~\eqref{effective polaron Hamiltonian} and leading to carrier localization, due to multiple scattering, at timescales of around 100 fs to 200 fs in agreement with the timescale of transient localization \cite{Fratini2016,Hutsch2022} in the absence of magnetic fields.  This dependence suggests that the transport states contributing to the channel mobility become localized for intermediate times $\tau$ and only the more extended transport states survive at larger times. 
In stark contrast, these localization effects for $\mu$ are not found in $\mu_{\text{H}}$. This means that vibrational disorder has a smaller effect on those transport states that respond to the magnetic field by deflection, which is in full agreement with experimental findings \cite{PodHall2005} (see below).

Furthermore, we show in Figs.~\ref{fig:fig2}(b) and (d) that the mobilities for a fixed value of $\tau$ do not vary significantly with the carrier concentration $n$  at least for $n\leq10^{-2}$ that are achievable experimentally. That is, the transport characteristics are quite homogeneous at the top of the HOMO band and therefore should not depend much on the gate voltage.   

\begin{figure}[t!]
	\centering
	\includegraphics[width=0.95\linewidth]{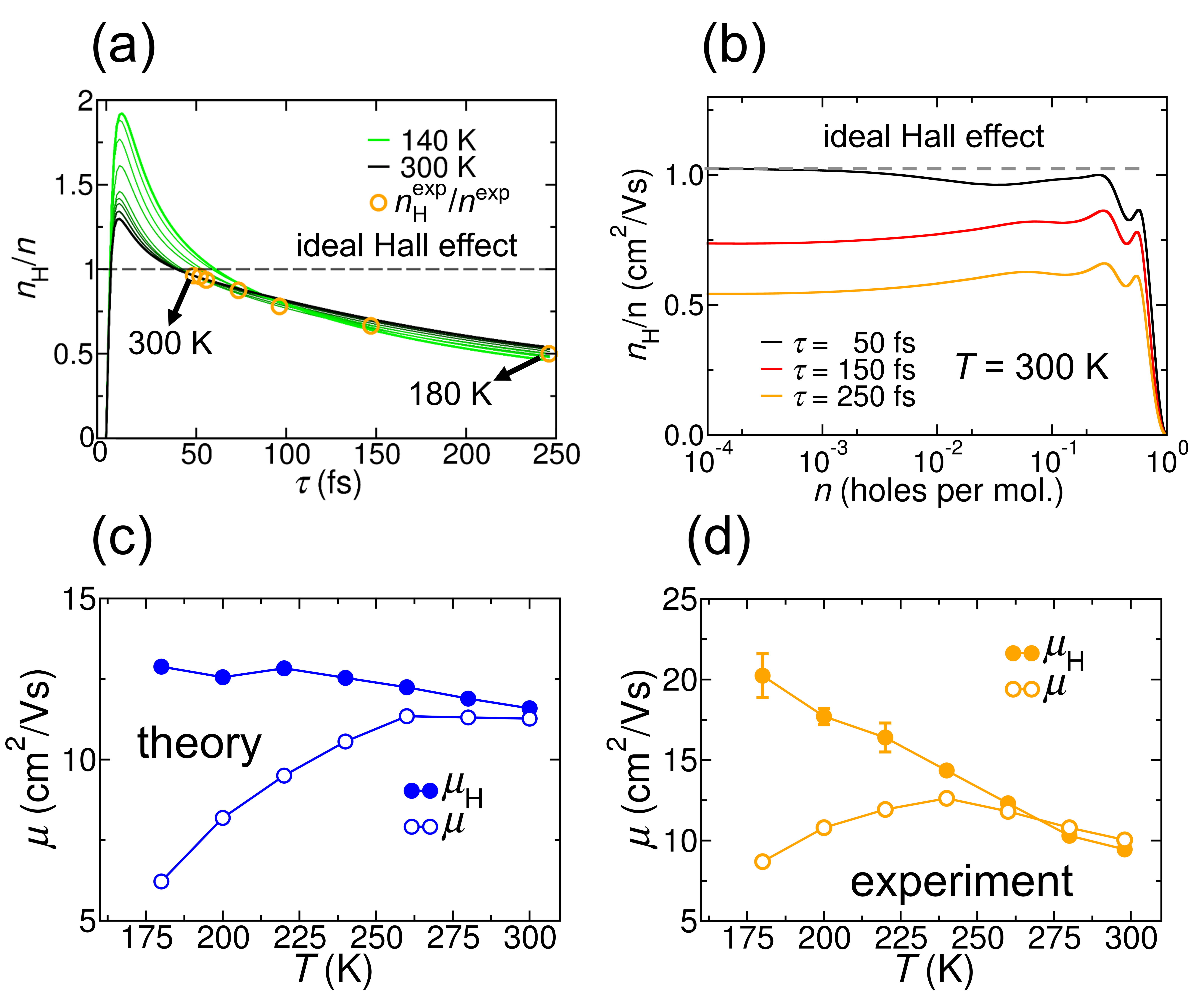}
	\caption{Temperature-dependent Hall transport in bulk rubrene crystals at 6 T. (a) Time dependence of the Hall density for temperatures between 140 K and 300 K in steps of 20 K.
		(b) Concentration dependence of the Hall density for fixed $\tau$ of 50 fs, 150 fs, and 250 fs at 300K. In (a) the orange circles indicate the experimental reference values obtained in Ref.~\cite{PodHall2005}. The gray dashed lines in (a) and (b) indicate the equality between the carrier density and the Hall density $n=n_\text{H}$.  (c) Theoretical Hall and channel mobilities. (d) Experimental Hall and channel mobilities taken from Ref.~\cite{PodHall2005}. }
	\label{fig:fig3}
\end{figure}

To finally compare our results, characterized by a distinct time and concentration dependence, to experimental data, for which this information is not directly available, one needs to fix certain values for $\tau$ and $n$. 
For the carrier concentration, we use $n=2.5\times10^{-4}$ holes per molecule, which is similar to the experimental value of roughly 10$^{10}$ cm$^{-2}$ and which has already been used to calculate the mobilities shown in Figs.~\ref{fig:fig2}(a) and (c).
To fix the relaxation time $\tau$, we match the numerically calculated value of the ratio $\mu/\mu_\text{H}\equiv n_\text{H}/n$ with the experimental ratio $\mu^\text{exp}/\mu_\text{H}^\text{exp}\equiv n_\text{H}^\text{exp}/n^\text{exp}$ of Ref.~\cite{PodHall2005} (cf. Fig.~\ref{fig:fig3}(d)).  
This is possible because of the consistency of experimental and theoretical results. More specifically, we find in Fig.~\ref{fig:fig3}(a) that $n_\text{H}/n$ also decays over time at fixed $n$ and is almost constant in the entire HOMO band at fixed $\tau$ (see Fig.~\ref{fig:fig3}(b)). Therefore, equating the mobility ratios yields the relaxation times as indicated by circles in Fig.~\ref{fig:fig3}(a).
We find that the resulting transport time scales, ranging from 50 fs at 300 K to 250 fs at 180 K, are consistent with the transport time scale within transient localization theory. \cite{Fratini2016}
In contrast to transient localization theory, in which $\tau$ is related to characteristic time scales of low-energy vibrations, the value of $\tau$ is here determined from the comparison of the experimental and the theoretical Hall density, fixing $\tau$ as a free parameter. We emphasize that this is not a contradiction to transient localization theory since the present values of $\tau$ are smaller than the typical time scale of low-energy vibration modes, which is the upper limit of the time scale where the vibrational-disorder model is a valid treatment of the EPC.

Based on the fixed values for $n$ and $\tau$, we calculate the temperature-dependent mobilities shown in Fig.~\ref{fig:fig3}(c). For comparison, Fig.~\ref{fig:fig3}(d) shows the experimental Hall measurement carried out by Podzorov \textit{et al.} \cite{PodHall2005} on vacuum-gated rubrene single-crystal OFETs.
Both theoretical and experimental results agree in that Hall and channel mobilities are equal at room temperature but differ at lower temperatures according to $\mu_\text{H}>\mu$. This indicates the formation of localized carriers due to vibrational disorder at low $T$. 
We note that the temperature-dependence of the channel mobility $\mu$ in Fig.~\ref{fig:fig3}(c)  especially its downturn towards lower temperatures (with $d\mu(T)/dT>0$) is a direct consequence of the treatment of the EPC as vibrational disorder within the transient localization approach. Given that the validity of transient localization is limited to transport time scales of up to a few hundred femtoseconds and may not extend to the lowest temperatures, we there also consider different predicted mobility behaviors. For instance, in diagrammatic Monte Carlo-based approaches without an assumed relaxation time, it was suggested that an activated temperature dependence ($d\mu(T)/dT>0$) at low temperatures only occurs at elevated EPC values.\cite{MishchenkoPRL2015,MishchenkoPRL2019}  This suggests a possible limitation of transient localization theory, at least for studied models that involve a single high-frequency vibration mode and a single charge carrier in a one-dimensional model crystal. However, a more detailed comparison is clearly beyond the scope of this work.
In general, the theoretical results are in good qualitative agreement to experimental mobility values, still we note that also static disorder caused, e.g., by surface disorder or impurities might lead to a mobility downturn due to carrier localization.

We note that the difference between the channel and Hall mobilities has been explained previously with a phenomenological model put forward by Yi \textit{et al.} \cite{Pod20162Hall}. In this model, the difference is attributed to the presence of two distinct types of charge carriers: band-like carriers and hopping carriers, the latter of which do not respond to the magnetic field.
Different to this minimal semi-classical model, in our approach all types of transport states and all of their potential coherent interferences are naturally included. An attempt was made to apply such phenomenological models for illustration, but the results were not satisfactory, which might be due to the absence of clearly distinct types of carriers or the effect of multiple-scattering and localization.

\begin{figure}[t!]
\centering
\includegraphics[width=0.95\linewidth]{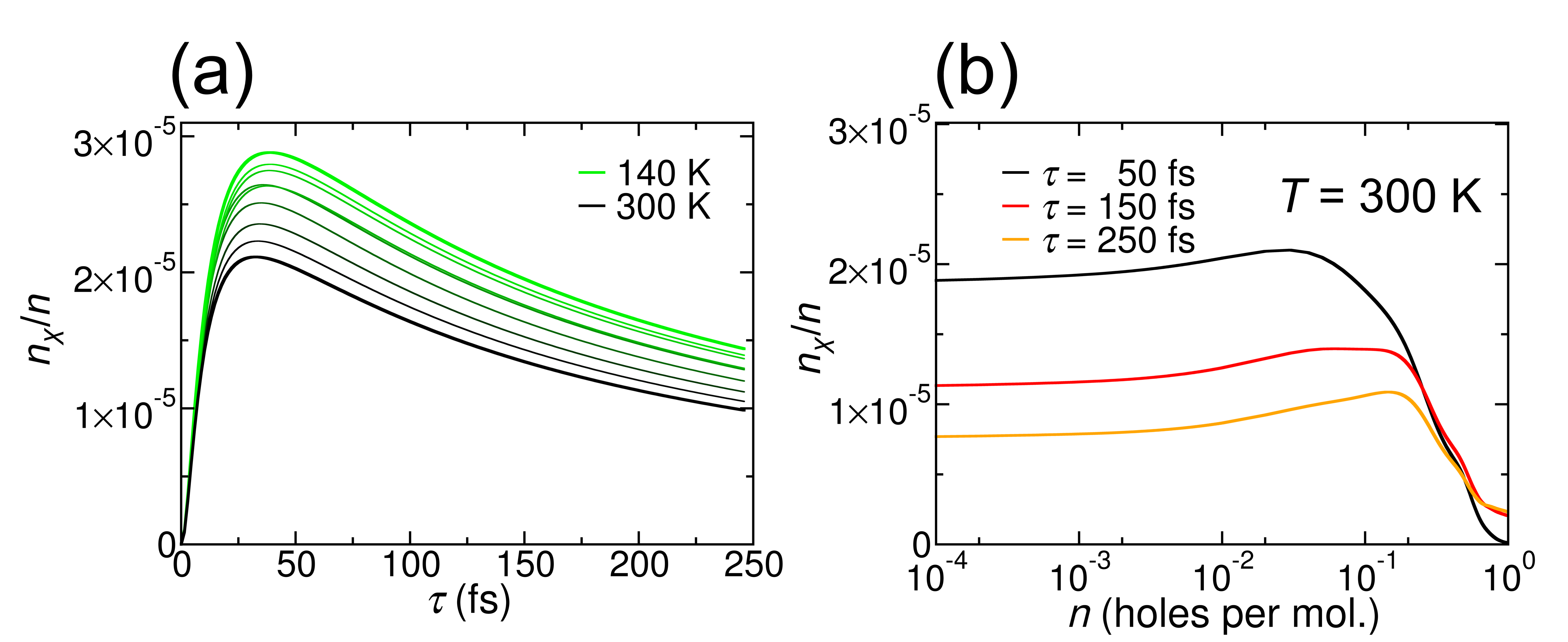}
\caption{Time and concentration dependence of $n_\chi$. (a): Time dependence of the density $n_\chi$ for temperatures between 140 K and 300 K in steps of 20 K.
	(b): Concentration dependence of the density $n_\chi$ for fixed $\tau$ of 50 fs, 150 fs, and 250 fs at 300K.
}
\label{fig:fig4}
\end{figure}
Instead, the understanding of the Hall transport in rubrene can be enhanced by analyzing the density $n_\chi=\sigma_\text{H}B_\gamma$ defined in Eq.~\eqref{en_chi} and connect the present Hall-transport regime to the intrinsic disorder-free case discussed above (Sect.~\ref{subsec:intrinsic}).
Specifically, for $n_\chi$ we find similar dependencies on the carrier concentration $n$ and on $\tau$ with its decay over time (cf. Fig.~\ref{fig:fig4}).
 For a fixed value of $\tau$, the ratio $n_\chi/n$ changes only weakly in the entire HOMO band, which is similar to $\mu_\text{H}$, $\mu$, and $n_\text{H}/n$. This result indicates that the number of states that respond to the magnetic field also increases uniformly with $n$ for carrier concentrations up to $10^{-1}$.
The difference to the intrinsic case discussed earlier, however, is very pronounced and eventually quite insightful.
While the Hall density $n_\text{H}$, both in the disorder-free and in the disordered case, is of the order of the carrier concentration $n_\text{H}\approx n$ (see above), the density $n_\chi$ is now five orders of magnitude smaller than $n$. This behavior is totally different to the pristine case, where we found $n_\chi \approx n$ for $n\leq$ 0.28 holes per molecule, which we attributed to the distribution of hole-like, electron-like, and open orbits in the HOMO band of rubrene (see discussion of Fig.~\ref{fig:fig1}(c)). In stark contrast here, the vibrational disorder leads to a breakdown of this semi-classical orbital picture since $n_\chi /n\approx 10^{-5}$ is much smaller.

Moreover, this behaviour indicates the breakdown of any semi-classical transport picture as a consequence of multiple scattering at the vibrational disorder.
Still, even in this regime, we find an ideal Hall effect ($n=n_\text{H}$) at around room temperature and only moderate non-ideality factors of about 0.5 when reducing temperature below 200 K. This ideal Hall effect in the transient localization regime is therefore distinguished from the semi-classical one in the intrinsic regime.

From these results we conclude that the presence of vibrational disorder (or disorder in general) has two significant effects: (i) The decay of $\mu$ as well as $n_\text{H}/n$ and $n_\chi/n$ with the relaxation time $\tau$, indicating the presence of localized states, which do not spread after having reached their typical localization length. This reduces the absolute value of $n_\chi$ yielding a net value of orders of magnitudes smaller than in the pristine case.
(ii) The difference between longitudinal and transverse response in presence of disorder leads to non-idealities with $n_\text{H}<n$.
(iii) Sufficiently large disorder can further lead to the intermixing, i.e., the hybridization of energetically separated hole-like, electron-like, and open-orbit energy domains, which can also lead to a breakdown of the ideal Hall effect with a different characteristic.

Notwithstanding the detailed analysis of the present Hall-transport regime, no indication of possible quantization effects in the transport characteristics were found. 
Beyond the straightforward observation that the temperatures considered in Sect.~\ref{subsec:temperature-Hall} are generally too high and magnetic fields possibly too low to observe quantization effects, there are primarily other reasons for their absence, prevailing even at sufficiently low $T$ and high $B_\gamma$.  
Fortunately, the proposed theory facilitates exploration of the transport regime, where the IQHE could emerge in organic semiconductors, and we wish to illustrate the conditions for the case of rubrene in the following section.

\subsection{\uppercase{IQHE in rubrene monolayers with disorder}}\label{subsec: IQHE}
\begin{figure}[t!]
	\centering
	\includegraphics[width=1\linewidth]{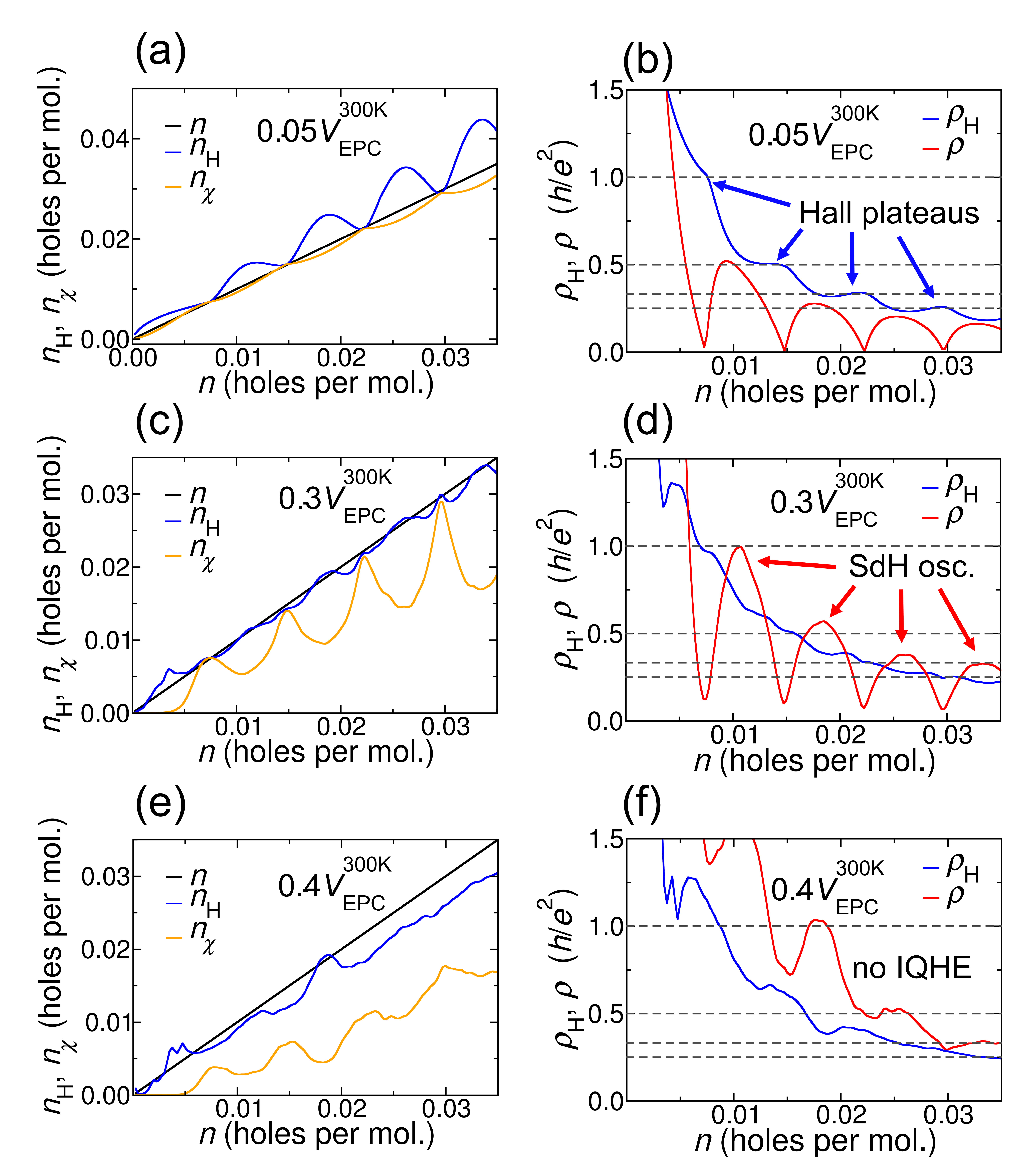}
	\caption{IQHE in a rubrene monolayer at 60 T for reduced disorder strengths. (a),(c),(e) Hall density $n_\text{H}$ and $n_\chi$ as a function of the charge-carrier concentration for disorder strengths of 2.9 meV, 17.4 meV, and 23.0 meV. The carrier concentration $n$ (solid black line) is also plotted for comparison. (b),(d),(f) Resistivity components $\rho$ and $\rho_\text{H}$ as a function of the total charge-carrier concentration for disorder strengths of 2.9 meV, 17.4 meV, and 23.0 meV. The dashed horizontal lines are the universal plateau heights of the IQHE regime.}
	\label{fig:fig5}
\end{figure}

We finally study the conditions for the emergence of the IQHE in rubrene monolayers with herringbone geometry in the presence of weak disorder. This scenario is suggested by transistor geometries where Hall transport occurs at the crystal surface in contact with the gate dielectric, leading to an ultrathin transport channel. For instance hexagonal boron nitride may be used as dielectric layer to generate large-area crystal domains \cite{Lee2014a}. For the observation of the IQHE, a certain amount of disorder is needed to generate localized states near the edges of the Landau bands, leading to a vanishing channel resistance $\rho$ and a quantized Hall resistance $\rho_{\text{H}}$ of $\rho_{\text{H}}=h/(e^2 \nu)$ with $\nu$ being an integer \cite{Klitzing1986}. 
To model single herringbone planes of rubrene, we simply disregard the small electronic coupling perpendicular to the herringbone plane. Furthermore, we set the temperature to 0 K and increase $B_\gamma$ to 60 T to prepare suitable IQHE conditions, which can be realized in high-magnetic field labs \cite{Zhu2017}. Vibrational disorder is known to exist in rubrene at room temperature and we determine its value to $V_\text{EPC}^{300\text{K}}=57.5$ meV (cf. Appendix \ref{app:F}). Taking this as reference value, we study different strengths of reduced disorder (relative to $V_\text{EPC}^{300\text{K}}$) in the following.
Fig.~\ref{fig:fig5} compiles the calculated $en_\text{H}$, $en_\chi$, and the resistivities $\rho$ ($\alpha$ direction) and $\rho_{\text{H}}$. 
We start in Figs.~\ref{fig:fig5}(a) and (b) discussing very small disorder strengths of 2.9 meV, which is below the spacing of the Landau-band centers ($\approx$5 meV). This value indeed leads to resistivity quantization in Fig.~\ref{fig:fig5}(b). Fig.~\ref{fig:fig5}(a) shows that the Hall density $n_\text{H}$ equals $n_\chi$ at those carrier concentrations at which the Hall plateaus appear in $\rho_{\text{H}}$, while $\rho$ vanishes in Fig.~\ref{fig:fig5}(b). The plateaus appear exactly with the quantized values of $h/(e^2 \nu)$ which is the famous IQHE. The regions of vanishing channel resistance $\rho$ are accompanied by pronounced SdH oscillations, indicating the presence of localized states. 

The certainly stronger disorder in experiments can obscure this quantization of $\rho_\text{H}$ under the chosen conditions for temperature and magnetic-field. Therefore, assessing the stability of the IQHE against a stronger disorder is critical. When the disorder strength is increased to 30\% of $V_\text{EPC}^{300\text{K}}$ (17.4 meV), the previously distinct plateaus get smeared out (Fig.~\ref{fig:fig5}(d)), which coincides with the vanishing of the step structure of $n_\text{H}$ plotted in Fig.~\ref{fig:fig5}(c).
The density $n_\chi$ only shows oscillations with varying carrier concentration indicating the pronounced intermixing of Landau bands due to the larger disorder. Despite the absence of clean Hall plateaus, we still observe clean SdH oscillations in $\rho$ in Fig.~\ref{fig:fig5}(d). Further increasing the disorder to 40\% of $V_\text{EPC}^{300\text{K}}$ (23.0 meV) leads to a breakdown of the IQHE. Even the SdH oscillations, which are generally more resilient to stronger disorder, begin to diminish (see Fig.~\ref{fig:fig5}(f)). Synchronously, the Hall density in Fig.~\ref{fig:fig5}(e) starts deviating from the charge-carrier density ($n_\text{H}<n$), indicating that the Hall effect becomes non-ideal. We explain this by the reduced localization length of transport states and the intermixing of Landau bands with different Chern numbers \cite{ShengPRL1997,YangPRL1996,LiuPRL1996} caused by the increasing amount of disorder. The intermixing of the Landau bands is a consequence of the finite bandwidth of the HOMO of rubrene leading to a complex superposition of different transport contributions. This reasoning for the transport regime of the IQHE is fully consistent with our interpretation of the Hall-transport regime for bulk crystals in Sect.~\ref{subsec:temperature-Hall}. 
More explicitly, we conclude that the intermixing of Landau bands with different Chern numbers is similar to the intermixing of hole-like, electron-like, and open-orbit energy domains as described above. We conclude that the intermixing of Landau bands and the intermixing of the three types of orbits is particularly important for OSCs since their typical electronic bandwidths remain usually below several hundreds of meV and \textit{must} include both Landau bands with different Chern numbers as well as electron-like and hole-like energy domains in the band structure in full agreement with the results for rubrene (cf. Figs.~\ref{fig:fig1} , \ref{fig:fig4}, and \ref{fig:fig5}).
\section{Summary and Conclusion}
In summary, this theoretical framework allows covering a variety of transport regimes, the concept of closed quasiparticle orbits in the BZ, temperature-dependent Hall transport in presence of vibrational disorder, and topological quantum effects such as the IQHE. 
The regimes have been characterized by distinctive relationships between different densities and distinct mobility responses.
The approach allows rationalizing in detail available experimental data and can be readily applied to the wide class of organic semiconducting materials. This will also support the interpretation of Hall measurements in the future and we hope to inspire further research in this direction.

\begin{acknowledgements}
	We would like to thank the Deutsche Forschungsgemeinschaft for financial support [projects No. OR-349/3 and OR-349/11 and the Clusters of Excellence e-conversion (Grant No. EXC2089) and MCQST (Grant No. EXC2111)]. Grants for computer time from the Leibniz Supercomputing Centre in Garching are gratefully acknowledged.
\end{acknowledgements} 

%

\newpage
\onecolumngrid
\appendix
\section{\uppercase{Derivation of the dc-conductivity}}\label{app:A}

In this appendix and in Appendix \ref{app:B}, we derive Eq.~\eqref{DC-conductivity} of the main text based on an efficient time-domain approach of the Kubo formalism.~\cite{Panhans2021PRL} For this purpose, we need to consider the general Kubo formula for the dc-conductivity $\sigma_{\alpha\beta}$ based on the linear response of the electric current density $J_\alpha=\text{Tr}\left(\hat{\rho}\hat{\jmath}_\alpha\right)$ to a constant electric field $E_\beta$ \cite{Kubo1957},
\begin{align}
\sigma_{\alpha\beta}=\frac{e^2}{V}\int\limits_{0}^\infty dt\,f_{v_\alpha x_\beta}(t), \label{general Kubo}
\end{align}
with the linear-response function $f_{v_\alpha x_\beta}(t)$.
We may express $f_{v_\alpha x_\beta}(t)$ in terms of spatial displacement operators $\Delta\hat{x}_\alpha(t)=\hat{x}_\alpha(t)-\hat{x}_\alpha(0)$ as derived previously \cite{Panhans2021PRL}:
\begin{align}
\begin{aligned}
f_{v_\alpha x_\beta}(t)&=\frac{1}{2\hbar}\frac{d}{dt}\left[\tan\left(\frac{\beta\hbar}{2}\frac{d}{dt}\right)\mathcal{D}^+_{x_\alpha x_\beta}(t)+\mathcal{D}^-_{x_\alpha x_\beta}(t)\right],\label{linear response function}
\end{aligned}
\end{align}
where the displacement-operator anticommutator function (DAF) is defined as
\begin{align}
\mathcal{D}^+_{x_\alpha x_\beta}(t)&=\text{Tr}\left(\hat{\rho}_0\left\{\Delta\hat{x}_\alpha(t),\Delta\hat{x}_\beta(t)\right\}\right) \label{DAF appendix}
\end{align}
and the displacement-operator commutator function (DCF) is defined as
\begin{align}
\mathcal{D}^-_{x_\alpha x_\beta}(t)&=-i\,\text{Tr}\left(\hat{\rho}_0\left[\Delta\hat{x}_\alpha(t),\Delta\hat{x}_\beta(t)\right]\right).\label{DCF appendix}
\end{align}
The square brackets and the curly brackets denote the commutator and the anticommutator of the spatial displacements, respectively. 
The time evolution of the involved operators is governed by the Hamiltonian $\hat{H}$, while for the thermal average the grand-canonical equilibrium density operator $\hat{\rho}_0=e^{-\beta\left(\hat{H}-\mu\hat{N}\right)}/\text{Tr}(e^{-\beta\left(\hat{H}-\mu\hat{N}\right)})$ is used. 
We note that the initial values of the DAF and the DCF vanish, i.e., $\mathcal{D}^\pm_{x_\alpha x_\beta}(0)=0$.

As a first step of the derivation of Eq.~\eqref{DC-conductivity}, we prove that Eqs.~\eqref{general Kubo} to \eqref{DCF appendix} can be written as
\begin{equation}
\begin{aligned}
\sigma_{\alpha\beta}=\frac{e^2}{2V}\lim\limits_{t\to\infty}\left[\frac{\beta}{2}\frac{d}{dt}\mathcal{D}^+_{x_\alpha x_\beta}(t)+\frac{1}{\hbar}\mathcal{D}^-_{x_\alpha x_\beta}(t)\right]. \label{dc-conductivity}
\end{aligned}
\end{equation}
Towards this end, the second term in Eq.~\eqref{dc-conductivity}, is readily obtained via time integration,
\begin{align}
\frac{e^2}{2\hbar V}\int_0^\infty dt\,\frac{d}{dt}\mathcal{D}^-_{x_\alpha x_\beta}(t)=\frac{e^2}{2\hbar V}\lim\limits_{t\to\infty}\mathcal{D}^-_{x_\alpha x_\beta}(t).
\end{align}
The first term of Eq.~\eqref{DC-conductivity} is derived from the first term in Eq.~\eqref{linear response function} and requires more steps. It can be identified with the time-symmetric part of the linear response function
\begin{align}
f_{v_\alpha x_\beta}^{\text{ts}}(t)=\frac{1}{2\hbar}\tan\left(\frac{\beta\hbar}{2}\frac{d}{dt}\right)\frac{d}{dt}\mathcal{D}^+_{x_\alpha x_\beta}(t).
\label{fts}
\end{align}
Next, we use that the second derivative of the DAF with respect to $t$ can be expressed with the time-symmetric part of the current-current correlation function
\begin{align}
\mathcal{S}_{v_\alpha v_\beta}(t)=\frac{1}{2}\text{Tr}\left(\hat{\rho}_0\left\{\hat{v}_\beta(0),\hat{v}_\alpha(t)\right\}\right)
\end{align}
according to
\begin{align}
\mathcal{S}^{\text{ts}}_{v_\alpha v_\beta}(t)=\frac{1}{4}\frac{d^2}{dt^2}\mathcal{D}^+_{x_\alpha x_\beta}(t).
\end{align}
Its substitution into Eq.~\eqref{fts} yields
\begin{align}
f_{v_\alpha x_\beta}^{\text{ts}}(t)
=\beta\frac{\tan\left(\frac{\beta\hbar}{2}\frac{d}{dt}\right)}{\frac{\beta\hbar}{2}\frac{d}{dt}}\mathcal{S}^\text{ts}_{v_\alpha v_\beta}(t).
\end{align}
As a next step, we express $\mathcal{S}^{\text{ts}}_{v_\alpha v_\beta}(t)$ by its Fourier transform and obtain
\begin{align}
f_{v_\alpha x_\beta}^{\text{ts}}(t)=\beta\frac{\tan\left(\frac{\beta\hbar}{2}\frac{d}{dt}\right)}{\frac{\beta\hbar}{2}\frac{d}{dt}}\left(\frac{1}{2\pi}\int\limits_{-\infty}^\infty d\omega\,e^{i\omega t}\mathcal{S}^{\text{ts}}_{v_\alpha v_\beta}(\omega)\right),
\end{align}
which can be simplified by applying the differential operator onto the exponential function, yielding
\begin{align}
f_{v_\alpha x_\beta}^{\text{ts}}(t)=\frac{\beta}{2\pi}\int\limits_{-\infty}^\infty d\omega\,e^{i\omega t}\frac{\tanh\left(\frac{\beta\hbar\omega}{2}\right)}{\frac{\beta\hbar\omega}{2}}\mathcal{S}^{\text{ts}}_{v_\alpha v_\beta}(\omega).
\end{align}
Calculating the time integral results in
\begin{align}
\begin{aligned}
\int\limits_0^\infty dt\,f_{v_\alpha x_\beta}^{\text{ts}}(t)
&=\frac{\beta}{2\pi}\int\limits_{-\infty}^\infty d\omega\,\left(\pi\delta(\omega)+\frac{i}{\omega}\right)\frac{\tanh\left(\frac{\beta\hbar\omega}{2}\right)}{\frac{\beta\hbar\omega}{2}}\mathcal{S}^{\text{ts}}_{v_\alpha v_\beta}(\omega).\label{time intergal time-symmetric response}
\end{aligned}
\end{align}
Since the function $\mathcal{S}^\text{ts}_{v_\alpha v_\beta}(t)$ is real and time-symmetric, its Fourier transform $\mathcal{S}^\text{ts}_{v_\alpha v_\beta}(\omega)$ is $\omega$-symmetric. Due to this symmetry, the second term on the right-hand side of Eq.~\eqref{time intergal time-symmetric response} vanishes because an odd function is integrated over a symmetric interval. Consequently, we remain with the first term given by
\begin{align}
\begin{aligned}
\int\limits_0^\infty dt\,f_{v_\alpha x_\beta}^{\text{ts}}(t)
&=\frac{\beta}{2}\mathcal{S}^{\text{ts}}_{v_\alpha v_\beta}(\omega=0)=\beta\int_0^\infty\mathcal{S}^{\text{ts}}_{v_\alpha v_\beta}(t)=\frac{\beta}{4}\int_0^\infty dt\,\frac{d^2}{dt^2}\mathcal{D}^+_{x_\alpha x_\beta}(t).
\end{aligned}
\end{align}
This identity can be integrated over time yielding the desired form of the first term of Eq.~\eqref{dc-conductivity}:
\begin{align}
\frac{\beta}{4}\int_0^\infty dt\,\frac{d^2}{dt^2}\mathcal{D}^+_{x_\alpha x_\beta}(t)=\frac{\beta}{4}\lim\limits_{t\to\infty}\frac{d}{dt}\mathcal{D}^+_{x_\alpha x_\beta}(t), \label{first term dc-conductivity}
\end{align}
because also the time derivative of $\mathcal{D}^+_{x_\alpha x_\beta}(t)$ vanishes at zero time.
We thus have shown that the dc-limit of the electrical conductivity is given by Eq.~\eqref{dc-conductivity} in its general from.

We proceed by introducing a finite relaxation time $\tau$ and rewrite Eq.~\eqref{dc-conductivity} as
\begin{align}
\sigma_{\alpha\beta}&=\lim\limits_{\tau\to\infty}\sigma_{\alpha\beta}(\tau),
\end{align}
with
\begin{align}
\begin{aligned}
\sigma_{\alpha\beta}(\tau)=\frac{e^2}{2V}\int_0^\infty dt\,e^{-\frac{t}{\tau}}\left[\frac{\beta}{2}\frac{d^2}{dt^2}\mathcal{D}^+_{x_\alpha x_\beta}(t)+\frac{1}{\hbar}\frac{d}{dt}\mathcal{D}^-_{x_\alpha x_\beta}(t)\right],
\end{aligned}
\end{align}
which we evaluate explicitly for non-interacting quasiparticles in Apppendix \ref{app:B} to obtain Eq.~\eqref{DC-conductivity}.

\section{\uppercase{ Derivation of the dc-conductivity for independent quasiparticles}}\label{app:B}
To obtain Eq.~\eqref{DC-conductivity} of the main text, we assume quasi-non-interacting (fermionic) quasiparticles, which applies if the Hamiltonian of the system has the common form
\begin{align}
\hat{H}_0=\sum\limits_{k}\varepsilon_k\hat{c}_k^\dagger\hat{c}_k,
\end{align}
with the eigenenergies $\varepsilon_k$ and the fermionic creation and annihilation operators $\hat{c}_k^\dagger$ and $\hat{c}_k$. Based on this assumption, we can further simplify the many-body thermal averages in the DAF and the DCF by applying Wick's theorem \cite{DANIELEWICZ1984239}. In general, we can write the electric dipole operator in terms of the fermionic eigenmodes as
\begin{align}
e\hat{x}_\alpha=e\sum\limits_{kl}x^\alpha_{kl}\hat{c}^\dagger_k\hat{c}_l.
\end{align}
The corresponding displacement operator is then simply evaluated as
\begin{align}
\Delta\hat{x}_{\alpha}(t)=\sum\limits_{kl}\left(e^{\frac{it}{\hbar}\left(\varepsilon_k-\varepsilon_l\right)}-1\right)x^\alpha_{kl}\hat{c}^\dagger_k\hat{c}_l=\sum\limits_{kl}\Delta x^\alpha_{kl}(t)\hat{c}^\dagger_k\hat{c}_l.
\end{align}
For the DAF, we thus find
\begin{equation}
\begin{aligned}
\mathcal{D}^+_{x_\alpha x_\beta}(t)
&=\int\limits_{-\infty}^\infty  dE\int\limits_{-\infty}^\infty  dE'f(E')(1-f(E))
\text{Tr}\left[\delta(E-\hat{H})\Delta\hat{x}_\alpha(t)\delta(E'-\hat{H})\Delta\hat{x}_\beta(t)+h.c. \right],
\end{aligned} 
\end{equation}
where we have introduced the spectral projection operator $\delta(E-\hat{H})$ and the completeness relation $\mathds{1}=\int\limits_{-\infty}^\infty dE\delta(E-\hat{H})$ with respect to the one-particle spectrum of the Hamiltonian. For the second derivative of the DAF, we obtain
\begin{equation}
\begin{aligned}
\frac{d^2}{dt^2}\mathcal{D}^+_{x_\alpha x_\beta}(t)\
&=2\int\limits_{-\infty}^\infty  dE\int\limits_{-\infty}^\infty  dE'f(E')(1-f(E)) \cos\left(\frac{t(E-E')}{\hbar}\right)\\&\quad\times\text{Tr}\left[\delta(E-\hat{H})\hat{v}_\alpha(0)\delta(E'-\hat{H})\hat{v}_\beta(0)+h.c.\right].\label{DAF energybasis}
\end{aligned} 
\end{equation}
From this relation, we find
\begin{equation}
\begin{aligned}
\int\limits_0^\infty dte^{-\frac{t}{\tau}}\frac{d^2}{dt^2}\mathcal{D}^+_{x_\alpha x_\beta}(t)
&=2\hbar\pi\int\limits_{-\infty}^\infty  dE\int\limits_{-\infty}^\infty  dE'f(E')(1-f(E))\frac{1}{\pi}\frac{\frac{\hbar}{\tau}}{\frac{\hbar^2}{\tau^2}+(E-E')^2}\\
&\quad\times\text{Tr}\left[\delta(E-\hat{H})\hat{v}_\alpha(0)\delta(E'-\hat{H})\hat{v}_\beta(0)+h.c.\right]. 
\label{difference DAF1}
\end{aligned} 
\end{equation}
In this expression, we see that the energy integrals include a convolution with the Dirac series $\delta_\tau(E-E')=\hbar/(\pi\tau)/(\hbar^2/\tau^2+(E-E')^2)$ which corresponds to a match in the one-particle energies $E=E'$ in the limit $\tau\to\infty$.  
For the one-particle DAF, which is introduced in Eq.~\eqref{DAF}, we obtain
\begin{align}
\begin{aligned}
\int\limits_0^\infty dte^{-\frac{t}{\tau}}\frac{d^2}{dt^2}\mathscr{D}^+_{x_\alpha x_\beta}(E,t)
&=2\pi\hbar\int\limits_{-\infty}^{\infty}dE' \delta_\tau(E-E')\text{Tr}\left[\delta(E-\hat{H})\hat{v}_\alpha(0)\delta(E'-\hat{H})\hat{v}_\beta(0)+h.c.\right].
\label{difference DAF2}
\end{aligned}
\end{align}
Therefore, using Eqs.~\eqref{difference DAF1} and \eqref{difference DAF2}, we find
\begin{align}
\begin{aligned}
\beta\int\limits_0^\infty dte^{-\frac{t}{\tau}}\frac{d^2}{dt^2}\mathcal{D}^+_{x_\alpha x_\beta}(t)
&=-\int\limits_0^\infty dte^{-\frac{t}{\tau}}\int\limits_{-\infty}^\infty dE\,\frac{d f(E)}{d E}\frac{d^2}{dt^2}\mathscr{D}^+_{x_\alpha x_\beta}(E,t)\\
&\quad+\int\limits_0^\infty dt\int\limits_{-\infty}^\infty dE\int\limits_{-\infty}^\infty dE'\order{\delta_\tau(E-E')[f(E)-f(E')]}. \label{equivalence DAF DAF independent particles}
\end{aligned}
\end{align}
The above relation includes the expression used for the calculation of the symmetric part of the electrical conductivity tensor in Eq.~\eqref{DC-conductivity}. The residual term in Eq.~\eqref{equivalence DAF DAF independent particles} vanishes for degenerate states, i.e., for $E=E'$ and for $\hbar/\tau\ll E-E'$, i.e., if external relaxation mechanisms are considered small against the energy differences (as usually considered). Thus, we approximate the full DAF with the one-particle DAF as
\begin{align}
\begin{aligned}
\beta\int\limits_0^\infty dte^{-\frac{t}{\tau}}\frac{d^2}{dt^2}\mathcal{D}^+_{x_\alpha x_\beta}(t)
&\approx -\int\limits_0^\infty dte^{-\frac{t}{\tau}}\int\limits_{-\infty}^\infty dE\,\frac{df(E)}{d E}\frac{d^2}{dt^2}\mathscr{D}^+_{x_\alpha x_\beta}(E,t)
\end{aligned}
\end{align}
for any given relaxation time $\tau$. In the limit $\tau\to\infty$, we find the identity
\begin{align}
\beta\lim\limits_{t\to\infty}\frac{d}{dt}\mathcal{D}^+_{x_\alpha x_\beta}(t)=-\lim\limits_{\tau\to\infty}\int\limits_0^\infty dte^{-\frac{t}{\tau}}\int\limits_{-\infty}^\infty dE\,\frac{df(E)}{d E}\frac{d^2}{dt^2}\mathscr{D}^+_{x_\alpha x_\beta}(E,t)=- \int\limits_{-\infty}^\infty dE\,\frac{d f(E)}{d E}\lim\limits_{t\to\infty}\frac{d}{dt}\mathscr{D}^+_{x_\alpha x_\beta}(E,t). \label{DAF SI}
\end{align}
The DCF can be evaluated equivalently to Eq.~\eqref{DAF energybasis}:
\begin{align}
\begin{aligned}
\mathcal{D}^-_{x_\alpha x_\beta}(t)
&=\int\limits_{-\infty}^\infty dEf(E)\mathscr{D}^-_{x_\alpha x_\beta}(E,t),
\end{aligned} \label{DCF SI}
\end{align}
where $\mathscr{D}^-_{x_\alpha x_\beta}(E,t)$ is the one-particle DCF given in Eq.~\eqref{DCF}. Collecting the two expressions in Eqs.~\eqref{DAF SI} and \eqref{DCF SI} yields
\begin{align}
\begin{aligned}
\sigma_{\alpha\beta}
&=\lim\limits_{\tau\to\infty}\sigma_{\alpha\beta}(\tau),
\end{aligned}
\end{align}
with
\begin{align}
\begin{aligned}
\sigma_{\alpha\beta}(\tau)
&=\frac{e^2}{2V}\int\limits_{0}^\infty dt \,e^{-\frac{t}{\tau}}\int\limits_{-\infty}^{\infty}dE\left[-\frac{1}{2}\frac{d f(E)}{d E}\frac{d^2}{dt^2}\mathscr{D}^+_{x_\alpha x_\beta}(E,t)+\frac{1}{\hbar}f(E)\frac{d}{dt}\mathscr{D}^-_{x_\alpha x_\beta}(E,t)\right],
\end{aligned}
\end{align}
which is Eq.~\eqref{DC-conductivity}.

\section{\uppercase{Intrinsic Hall conductivity and orbit picture}}\label{app:C} 
In this section, we evaluate the intrinsic Hall conductivity in the absence of disorder and connect it to semi-classical expressions from semi-classical band theory. As a first step, we write the quasiparticle Hamiltonian without external potential but coupled to the operator for the vector potential $\mathbf{\hat{A}}$,
\begin{align}
\hat{H}=\hat{H}(\hat{\mathbf{p}}-q\hat{\mathbf{A}}). \label{Hamiltonian vector potential}
\end{align}
Here, we consider a homogeneous magnetic field $\mathbf{B}=\nabla\times\mathbf{A}$, which we may realize using the symmetric gauge for the operator of the vector potential
$\mathbf{\hat{A}}=\frac{1}{2}\mathbf{B}\times\mathbf{\hat{x}}$. 
We now set the magnetic field to $\mathbf{B}=B_\gamma\mathbf{e}_\gamma$, where $\mathbf{e}_\gamma$ is perpendicular to the transport directions $\mathbf{e}_\alpha$ and $\mathbf{e}_\beta$ yielding the vector potential $\mathbf{\hat{A}}=(-B_\gamma\hat{x}_\beta/2,B_\gamma\hat{x}_\alpha/2,0)$. We now introduce the kinetic momenta $\hat{\pi}_\alpha^\pm=\hat{p}_\alpha\pm q\hat{A}_\alpha$ and $\hat{\pi}_\beta^\pm=\hat{p}_\beta\pm q\hat{A}_\beta$. This transformation can be used to express both the position and canonical momentum operators with $\hat{\pi}_{\alpha/\beta}^\pm$ according to
\begin{align}
\begin{aligned}
\hat{x}_\alpha&=\frac{1}{qB_\gamma}\left(\hat{\pi}_\beta^+-\hat{\pi}_\beta^-\right),&&\hat{p}_\alpha=\frac{1}{2}\left(\hat{\pi}_\beta^++\hat{\pi}_\beta^-\right),\\
\hat{x}_\beta&=\frac{1}{qB_\gamma}\left(\hat{\pi}_\alpha^--\hat{\pi}_\alpha^+\right),&&\hat{p}_\beta=\frac{1}{2}\left(\hat{\pi}_\alpha^++\hat{\pi}_\alpha^-\right). \label{coordinate transformation}
\end{aligned}
\end{align} 
This turns out to be convenient since the Hamiltonian then solely depends on the kinetic momenta $\hat{\pi}_\alpha^-$ and $\hat{\pi}^-_\beta$.
The kinetic momentum operators satisfy the commutation relations
\begin{align}
\begin{aligned}
\left[\hat{\pi}_\alpha^-,\hat{\pi}_\beta^-\right]&=\left[\hat{\pi}_\beta^+,\hat{\pi}_\alpha^+\right]=i\hbar qB_\gamma,\\
\left[\hat{\pi}_\alpha^-,\hat{\pi}_\alpha^+\right]&=\left[\hat{\pi}_\beta^-,\hat{\pi}_\beta^+\right]=\left[\hat{\pi}_\alpha^-,\hat{\pi}_\beta^+\right]=\left[\hat{\pi}_\beta^-,\hat{\pi}_\alpha^+\right]=0.
\end{aligned}
\end{align} 
We now evaluate the Heisenberg equation of motion for the kinetic momentum operators:
\begin{align}
\begin{aligned}
\dot{\hat{\pi}}^+_\alpha(t)&=0,&&
\dot{\hat{\pi}}_\alpha^-(t)=qB_\gamma\frac{\partial \hat{H}}{\partial \hat{\pi}_\beta^-}(t),\\
\dot{\hat{\pi}}^+_\beta(t)&=0,&&
\dot{\hat{\pi}}_\beta^-(t)=-qB_\gamma\frac{\partial \hat{H}}{\partial \hat{\pi}_\alpha^-}(t).
\end{aligned}
\end{align}
and find a simple relation between the time derivatives of the position and the kinetic momentum operators
\begin{align}
\dot{\hat{\pi}}_\alpha^-(t)&=qB_\gamma\hat{v}_\beta(t),\\
\dot{\hat{\pi}}_\beta^-(t)&=-qB_\gamma\hat{v}_\alpha(t).
\end{align}

To evaluate the Hall conductivity $\sigma_\text{H}$ for this case, we decompose it into two terms
\begin{align}
\sigma_\text{H}=\sigma_\text{H}^\text{I}+\sigma_\text{H}^\text{II},
\end{align}
with 
\begin{align}
&\begin{aligned}
\sigma_\text{H}^\text{I}&=\frac{e^2}{2V}\int\limits_{-\infty}^\infty dE\,f(E)\text{Tr}\left[\delta(E-\hat{H})\left(\hat{v}_\alpha\left[\Re\hat{G}(E),\hat{x}_\beta\right]-\hat{v}_\beta\left[\Re\hat{G}(E),\hat{x}_\alpha\right]\right)\right],
\end{aligned}\\
&\sigma_\text{H}^\text{II}=\frac{e^2}{2V}\int\limits_{-\infty}^\infty dE\,f(E)\text{Tr}\left[\frac{d}{dE}\delta(E-\hat{H})\left(\hat{x}_\alpha\hat{v}_\beta-\hat{x}_\beta\hat{v}_\alpha\right)\right],
\end{align}
where $\Re\hat{G}(E)=(E-\hat{H})^{-1}$ is the real part of the Green's function. This decomposition of $\sigma_\text{H}$ is derived in detail in Appendices \ref{app:D} and \ref{app:E}.
We first calculate the second term of the Hall conductivity $\sigma_\text{H}^\text{II}$ by using the equations of motion for the kinetic momenta,
\begin{align}
\begin{aligned}
\text{Tr}\left[\delta(E-\hat{H})(\hat{v}_\alpha \hat{x}_\beta-\hat{v}_\beta\hat{x}_\alpha)\right]
&=\frac{1}{e^2B^2_\gamma}\text{Tr}\left[\delta(E-\hat{H})\left(\dot{\hat{\pi}}^-_\alpha\hat{\pi}_\beta^--\dot{\hat{\pi}}_\beta^- \hat{\pi}_\alpha^-\right)\right],
\end{aligned}
\end{align}
where we have used that the operators $\hat{\pi}_\alpha^+$ and $\hat{\pi}^+_\beta$ commute with the Hamiltonian.
We note that the expectation value only depends on the conjugate operators $\hat{\pi}_\alpha^-$ and $\hat{\pi}_\beta^-$. 

To obtain the semi-classical result for the Hall conductivity, we now substitute the quantum mechanical trace with the integral over the phase space
\begin{align}
\begin{aligned}
&\text{Tr}\left[\delta(E-\hat{H})(...)\right]\to\frac{1}{(2\pi\hbar)^2}\int dx_\alpha dx_\beta dp_\alpha dp_\beta\,\delta(E-\varepsilon(\pi_\alpha^-,\pi_\beta ^-))(...),
\end{aligned}
\end{align}
which we can substitute with the integral over the kinetic momenta using the coordinate transformation in Eq.~\eqref{coordinate transformation}:
\begin{align}
\begin{aligned}
&\frac{1}{(2\pi\hbar)^2}\int dx_\alpha dx_\beta dp_\alpha dp_\beta\,\delta(E-\varepsilon(\pi_\alpha^-,\pi_\beta ^-))(...)
\\&\quad=\frac{1}{(2\pi\hbar e B_\gamma)^2}\int d\pi_\alpha^+d\pi_\beta^+d\pi_\alpha^-d\pi_\beta^-\,\delta(E-\varepsilon(\pi_\alpha^-,\pi_\beta ^-))(...).
\end{aligned}
\end{align}
Since the total number of states is given by
\begin{align}
\begin{aligned}
N=\int\limits_{-\infty}^\infty dE\,\text{Tr}\left(\delta(E-\hat{H})\right)
\to \frac{1}{(2\pi\hbar eB_\gamma)^2}\int d\pi_\alpha^+d\pi_\beta^+ d\pi_\alpha^-d\pi_\beta^-
\end{aligned}
\end{align}
we obtain its semi-classical result as
\begin{align}
N=\frac{\Omega^2}{(2\pi\hbar eB_\gamma)^2},
\end{align}
with $\Omega$ being the phase-space volume of the kinetic momenta:
\begin{align}
\Omega=\int d\pi_\alpha^+d\pi_\beta^+=\int d\pi_\alpha^-d\pi_\beta^-.
\end{align}
The expectation value entering the Hall conductivity then reads
\begin{align}
\begin{aligned}
\text{Tr}\left[\delta(E-\hat{H})\left(\dot{\hat{\pi}}^-_\alpha\hat{\pi}_\beta^--\dot{\hat{\pi}}_\beta^- \hat{\pi}_\alpha^-\right)\right]
&=\frac{N}{\Omega}\int d\pi_\alpha^-d\pi_\beta^-\,\delta(E-\varepsilon(\pi_\alpha^-,\pi_\beta ^-))\left(\dot{\pi}_\alpha^- \pi_\beta^--\dot{\pi}^-_\beta\pi_\alpha^-\right),
\end{aligned}
\end{align}
where the integration over the kinetic momenta $\pi_\alpha^+$ and $\pi_\beta^+$ amounts to the phase-space volume $\Omega$ since the integrand is independent of these coordinates for the present model. We now may substitute the integral over the phase space of kinetic momenta with the integral along all quasiparticle orbits of constant energy (which represents the Fermi surface in the case of a 2D system):
\begin{align}
\begin{aligned}
\frac{N}{\Omega}\int d\pi_\alpha^-d\pi_\beta^-\,\delta(E-\varepsilon(\pi_\alpha^-,\pi_\beta ^-))\left(\dot{\pi}_\alpha^- \pi_\beta^--\dot{\pi}^-_\beta\pi_\alpha^-\right)
&=\frac{N}{\Omega}\int\limits_{\text{orbits}} d\boldsymbol{\pi}\,\frac{1}{\left|\nabla_{\boldsymbol{\pi}}\varepsilon(\pi_\alpha^-,\pi_\beta ^-)\right|}\left(\dot{\pi}_\alpha^- \pi_\beta^--\dot{\pi}^-_\beta\pi_\alpha^-\right).
\end{aligned}
\end{align} 
Using the equations of motion to substitute $\left|\nabla_{\boldsymbol{\pi}}\varepsilon(\pi_\alpha ^-,\pi_\beta^-)\right|=\left|\dot{\boldsymbol{\pi}}\right|/eB_\gamma$, we obtain
\begin{align}
\begin{aligned}
\frac{N}{\Omega}\int\limits_{\text{orbits}} d\pi\,\frac{1}{\left|\nabla_{\boldsymbol{\pi}}\varepsilon(\pi_\alpha^-,\pi_\beta ^-)\right|}\left(\dot{\pi}_\alpha^- \pi_\beta^--\dot{\pi}^-_\beta\pi_\alpha^-\right)&=\frac{N}{\Omega}\int \limits_{\text{orbits}}d\pi\,\frac{eB_\gamma}{\left|\dot{\boldsymbol{\pi}}\right|}\left[\dot{\pi}_\alpha^- \pi_\beta^--\dot{\pi}^-_\beta\pi_\alpha^-\right]\\
&=\frac{NeB_\gamma}{\Omega}\left(\int\limits_\text{orbits} d\boldsymbol{\pi}\,\times\boldsymbol{\pi}\right)_\gamma
\end{aligned}
\end{align}
The $\gamma$-component of the integral $\int_{\text{orbits}}d\boldsymbol{\pi}\times\boldsymbol{\pi}$ is related to the area $A_e(E)-A_h(E)$ enclosed by all semi-classical orbits with energy $E$ via
\begin{align}
\begin{aligned}
\left(\int_\text{orbits} d\boldsymbol{\pi}\,\times\boldsymbol{\pi}\right)_\gamma=2(A_e(E)-A_h(E)),
\end{aligned}
\end{align}
leading to
\begin{align}
\begin{aligned}
\text{Tr}\left[\delta(E-\hat{H})\left(\dot{\hat{\pi}}^-_\alpha\hat{\pi}_\beta^--\dot{\hat{\pi}}_\beta^- \hat{\pi}_\alpha^-\right)\right]
=2NeB_\gamma\frac{A_e(E)-A_h(E)}{\Omega}.
\end{aligned}
\end{align}
We now see that the expectation value is proportional to the difference of the areas $A_e(E)-A_h(E)$ covered by electron-like and hole-like states in the phase space, which are enclosed by all quasiparticle orbits with energy $E$.
Thus, we obtain the Hall conductivity $\sigma_\text{H}^\text{II}$ as
\begin{align}
\begin{aligned}
\sigma_\text{H}^\text{II}
&=\frac{eN}{B_\gamma V}\int\limits_{-\infty}^\infty dE\,f(E)\frac{d}{dE}\frac{A_e(E)-A_h(E)}{\Omega}.
\end{aligned}
\end{align}
In the same way, the remaining term $\sigma_\text{H}^\text{I}$ can be obtained using
\begin{align}
\begin{aligned}
&\text{Tr}\left[\delta(E-\hat{H})\left(\left[\hat{x}_\beta,\hat{v}_\alpha\Re\hat{G}(E)\right]-\left[\hat{x}_\alpha,\hat{v}_\beta\Re\hat{G}(E)\right]\right)\right]\\
&\quad=\frac{1}{e^2B_\gamma^2}\text{Tr}\left[\delta(E-\hat{H})\left(\dot{\hat{\pi}}^-_\alpha\left[\hat{\pi}_\beta^-,\Re\hat{G}(E)\right]+h.c.\right)\right].
\end{aligned}
\end{align}
The semi-classical result of this expectation value is obtained by substituting the commutator under the trace with the Poisson bracket:
\begin{align}
\left[...,...\right]\to i\hbar qB_\gamma\left\{...,...\right\}.
\end{align}
We find
\begin{align}
\begin{aligned}
\frac{1}{e^2B_\gamma^2}\text{Tr}\left[\delta(E-\hat{H})\left(\dot{\hat{\pi}}^-_\alpha\left[\hat{\pi}_\beta^-,\Re\hat{G}(E)\right]+h.c.\right)\right]
&=-\frac{iq\hbar N}{e\Omega}\int\limits_{\text{orbits}} d\boldsymbol{\pi}\,\cdot\nabla_{\boldsymbol{\pi}}\Re G(E)\\
&=0,
\end{aligned}
\end{align}
because the real part of the Green's function at energy $E$ is constant along all quasiparticle orbits.
Therefore, we obtain 
\begin{align}
\sigma_\text{H}^\text{I}=0,
\end{align}
which means that $\sigma_\text{H}^\text{I}$ vanishes identically in the semi-classical limit.
We thus obtain for the total intrinsic Hall conductivity as given in Eq.~\eqref{sigma_H sc}
\begin{align}
\sigma_\text{H}^\text{sc}=\sigma_\text{H}^\text{II}&=\frac{eN}{B_\gamma V}\int\limits_{-\infty}^\infty dE\,f(E)\frac{d}{dE}\frac{A_e(E)-A_h(E)}{\Omega_\text{BZ}},
\label{semi-classical result sigma_H}
\end{align}
where we have substituted the phase-space volume $\Omega$ with the volume of the Brillouin zone in the reciprocal space via $\Omega=\hbar^2\Omega_\text{BZ}$ and redefined $A_e(E)-A_h(E)$ with the area enclosed by the quasiparticle orbits in the reciprocal space. 
Based on the numerical results for the Hall conductivity, we may approximate the Hall conductivity solely in terms of the specific distribution of closed and open orbits associated with electron- and hole-like behavior. We have found the following intuitive expression for the Hall conductivity (at zero temperature)
\begin{align}
\begin{aligned}
\sigma_\text{H}(E_\text{F})&\approx -\frac{e}{B_\gamma}\left[n_e\Theta(E_\text{F}-E_e) -(1-n_e)(1-\Theta(E_\text{F}-E_h))\right],\label{sigma_H carrier density}
\end{aligned}
\end{align}
which agrees with Eq.~\eqref{approximate sigma_H} if we set the carrier density of the holes to $n=n_h=1-n_e$. The energies $E_e$ and $E_h$ correspond to the critical energies, up to which closed electron-like and hole-like orbits appear at the Fermi energy $E_\text{F}$. 

We now connect these general findings to the results for the intrinsic Hall conductivity of rubrene in Figs.~\ref{fig:fig1}(b) to \ref{fig:fig1}(d).
The energy dependence of $\sigma_\text{H}$ in Fig.~\ref{fig:fig1}(b) (or Fig.~\ref{fig:fig1}(c) for $en_\chi$) can be explained if we consider only the hole-like orbits with a Fermi energy larger than 103 meV (and for zero temperature). In this case, all quasiparticle orbits are closed and the area difference $A_e(E_\text{F})-A_h(E_\text{F})$ is equal to the area covered by the occupied hole states in the BZ $A_\text{occ}(E_\text{F})$, which is related to the carrier concentration $n=N_\text{occ}(E_\text{F})/V$ by $n/N=A_\text{occ}(E_\text{F})/\Omega_\text{BZ}$ with the number of occupied states $N_\text{occ}(E_\text{F})$. Then, from Eq.~\eqref{semi-classical result sigma_H} we immediately get $\sigma_\text{H}=qN/(B_\gamma V)A_\text{occ}(E_\text{F})/\Omega_\text{BZ}=en/B_\gamma$. This ideal behavior is perfectly satisfied in Figs.~\ref{fig:fig1}(b) and \ref{fig:fig1}(c) for Fermi energies above 103 meV, i.e., the hole-like states. 
If the Fermi energy lies between these values, the orbits are open in the BZ and the Hall conductivity becomes zero in qualitative agreement with the numerical result of Figs.~1(b) and 1(c). Then for Fermi energies below -202 meV a very small domain of electron-like orbits manifesting as a small negative peak in Figs.~\ref{fig:fig1}(b) and \ref{fig:fig1}(c), is observed in full agreement with the pristine band structure of rubrene.

We note that Eq.~\eqref{sigma_H carrier density} (i.e., Eq.~\eqref{approximate sigma_H}) is only an approximate realization of Eq.~\eqref{semi-classical result sigma_H} (i.e., Eq.~\eqref{sigma_H sc}) and may further depend on the details of the band structure in, e.g., three-dimensional bulk systems and if more complex band structures are considered, which may include multiple electron-like, hole-like, and open-orbit domains.

\section{\uppercase{Equilibrium value  of dc-conductivity tensor}}\label{app:D}
To prove the semi-classical expression for the Hall conductivity in Eq.~\eqref{sigma_H sc} of the main text, we need to determine the equilibrium value of the dc-conductivity based on Eq.~\eqref{DC-conductivity} for $\tau\to\infty$. After this derivation and in Appendix \ref{app:E}, we connect the present time-domain approach with known forms of the dc-conductivity derived within the Kubo formalism such as the Kubo-Bastin formula \cite{BASTIN19711811} and the Berry curvature that enters the TKNN formula \cite{Thouless1982PRL,NiuPRB1985}.
Hence, we evaluate the time integral in the symmetric part of the conductivity tensor $\sigma_{\alpha\beta}^\text{s}$ in Eq.~\eqref{symmetric part dc-conductivity}. In the energy eigenbasis, we find for $\sigma^\text{s}_{\alpha\beta}$:
\begin{align}
\begin{aligned}
\sigma_{\alpha\beta}^{\text{s}}
&=\frac{\beta e^2\pi\hbar}{2V}\sum\limits_{mn}f(\varepsilon_n)\left(1-f(\varepsilon_n)\right)\delta(\varepsilon_n-\varepsilon_m)\left(v^\alpha_{nm} v^\beta_{mn}+ v^\beta_{nm}v^\alpha_{mn}\right),
\end{aligned}
\end{align}
where we write the matrix elements $v^\alpha_{mn}=v^\alpha_{mn}(0)$ as a short-hand notation.
This expression is the equilibrium value for the symmetric part of the conductivity and generalizes the Kubo-Greenwood formula for longitudinal charge transport \cite{Greenwood_1958,Fan2021a}, which can be seen by using the completeness relation $\mathds{1}=\int_{-\infty}^\infty dE\delta(E-\hat{H})$:
\begin{align}
\begin{aligned}
\sigma_{\alpha\beta}^{\text{s}}
&=\frac{\beta e^2\pi\hbar}{V}\int\limits_{-\infty}^\infty dE\,f(E)(1-f(E))\text{Tr}\left(\delta(E-\hat{H})\hat{v}_\alpha\delta(E-\hat{H})\hat{v}_\beta\right).
\end{aligned}
\end{align}
The evaluation of the time integral in the Hall conductivity for independent particles (see Eq.~\eqref{anti-symmetric part dc-conductivity}) results in the following expression: 
\begin{align}
\begin{aligned}
\sigma_\text{H}
&=-\frac{e^2}{ V}\sum\limits_{mn}\frac{f(\varepsilon_n)}{\varepsilon_n-\varepsilon_m}\left(v^\alpha_{nm} x^\beta_{mn}-v^\beta_{nm}x^\alpha_{mn}\right),
\end{aligned}
\end{align}
where we have used the matrix elements of the velocity operator $v_{nm}^\alpha=i(\varepsilon_n-\varepsilon_m)x^\alpha_{nm}/\hbar$. Furthermore we may alternatively derive the Hall conductivity by using the following identity:
\begin{align}
\begin{aligned}\frac{d^2}{dt^2}\mathcal{D}^-_{x_\alpha x_\beta}(t)&=-2\hbar f_{v_\alpha v_\beta}(0)-\mathcal{D}^-_{v_\alpha v_\beta}(t)\\
&=-2i\sum\limits_{mn}f(\varepsilon_n)\left(v^\alpha_{nm} v^\beta_{mn}-v^\beta_{nm}v^\alpha_{mn}\right)\cos\left(\frac{t\left(\varepsilon_n-\varepsilon_m\right)}{\hbar}\right),
\end{aligned}
\end{align}
where $f_{v_\alpha v_\beta}(0)$ is the linear response function of the velocity operators at zero time \cite{Panhans2021PRL}. We then find for the Hall conductivity
\begin{align}
\begin{aligned}
\sigma_\text{H}
&=\frac{ie^2\hbar}{V}\lim\limits_{\tau\to\infty}\sum\limits_{mn}f(\varepsilon_n)\frac{\left(v^\alpha_{nm} v^\beta_{mn}-v^\beta_{nm}v^\alpha_{mn}\right)}{(\varepsilon_n-\varepsilon_m)^2+\left(\frac{\hbar}{\tau}\right)^2}.
\end{aligned}
\end{align}
The equilibrium value for the Hall conductivity for $\tau\to\infty$ is hence obtained as
\begin{align}
\begin{aligned}
\sigma_\text{H}&=\frac{ie^2\hbar}{V}\sum\limits_{mn}f(\varepsilon_n)\frac{\left(v^\alpha_{nm} v^\beta_{mn}-v^\beta_{nm}v^\alpha_{mn}\right)}{(\varepsilon_n-\varepsilon_m)^2}
,
\end{aligned}
\end{align}
which describes the thermally averaged Berry curvature \cite{Thouless1982PRL,NiuPRB1985}.
Using the spectral-projection operator, we obtain the Hall conductivity as
\begin{align}
\begin{aligned}
\sigma_\text{H}
&=\frac{ie^2\hbar}{V}\int\limits_{-\infty}^\infty dE\,f(E)\text{Tr}\left[\delta(E-\hat{H})\left(\hat{v}_\alpha\frac{1}{(E-\hat{H})^2} \hat{v}_\beta-h.c.\right)\right].
\end{aligned}
\end{align}
Thus, the equilibrium value of the dc-conductivity can equivalently be written as
\begin{align}
\begin{aligned}
\sigma_{\alpha\beta}&=-\frac{e^2\pi\hbar}{V}\int\limits_{-\infty}^\infty dE\,\frac{df(E)}{dE}\text{Tr}\left(\delta(E-\hat{H})\hat{v}_\alpha\delta(E-\hat{H})\hat{v}_\beta\right)
+\sigma_\text{H},
\end{aligned}\label{equilibrium value dc-conductivity}
\end{align}
which we derived directly from Eq.~\eqref{DC-conductivity}.

\section{\uppercase{Relation of the dc-conductivity to the Kubo-Bastin formula}}\label{app:E}
In this appendix, we show the equivalence of the dc-conductivity used in this work with the Kubo-Bastin formula \cite{BASTIN19711811} based on the equilibrium values of the symmetric and the antisymmetric parts of the dc-conductivity derived in Appendix \ref{app:D}. We start with the symmetric part of the conductivity and discuss its relation to the Kubo-Bastin formula. For this purpose, we introduce the Green's functions
\begin{align}
\hat{G}^\pm(E)=\lim\limits_{\eta\to 0}\frac{1}{E-\hat{H}\pm i\eta}=\lim\limits_{\tau\to \infty}\frac{1}{E-\hat{H}\pm \frac{i\hbar}{\tau}},
\end{align}
where the real and the imaginary parts are given by
\begin{align}
\Re\hat{G}(E)&=\frac{1}{E-\hat{H}}=\lim\limits_{\tau\to \infty}\frac{E-\hat{H}}{(E-\hat{H})^2 +\left(\frac{\hbar}{\tau}\right)^2}, \\\Im\hat{G}(E)&=-\pi\delta(E-\hat{H})=-\lim\limits_{\tau\to \infty}\frac{\frac{\hbar}{\tau}}{(E-\hat{H})^2+ \left(\frac{\hbar}{\tau}\right)^2}.
\end{align}
The Kubo-Bastin formula for the Hall conductivity reads \cite{BASTIN19711811}:
\begin{align}
\sigma_{\alpha\beta}^\text{Bastin}=\frac{e^2}{V}\int\limits_{-\infty}^\infty dE\,f(E)C_{\alpha\beta}(E),
\end{align}
with $f(E)$ being the Fermi function and $C_{\alpha\beta}(E)$ being the correlation function
\begin{align}
C_{\alpha\beta}(E)&=i\hbar\text{Tr}\left[\hat{v}_\alpha\delta(E-\hat{H})\hat{v}_\beta\frac{d\hat{G}^+(E)}{dE}-h.c.\right].
\end{align}

To prove the equivalence to the present version of the dc-conductivity based on Eq.~\eqref{DC-conductivity}, we may decompose $C_{\alpha\beta}(E)$ as follows:
\begin{align}
\begin{aligned}
C_{\alpha\beta}(E)
&=C_{\alpha\beta}^\text{s}(E)+C_{\alpha\beta}^\text{as}(E)\label{Kubo-Bastin I},
\end{aligned}
\end{align}

with $C_{\alpha\beta}^\text{s(as)}(E)=\left(C_{\alpha\beta}(E)+(-)C_{\beta\alpha}(E)\right)/2$ yielding

\begin{align}
&	\begin{aligned}
C_{\alpha\beta}^\text{s}(E)&=\pi\hbar\frac{d}{dE}\text{Tr}\left[\delta(E-\hat{H})\hat{v}_\alpha\delta(E-\hat{H})\hat{v}_\beta\right],
\end{aligned}
\\
&\begin{aligned}
C_{\alpha\beta}^\text{as}(E)&=i\hbar\text{Tr}\left[\delta(E-\hat{H})\left(\hat{v}_\alpha\frac{1}{(E-\hat{H})^2}\hat{v}_\beta-h.c.\right)\right],
\end{aligned}
\end{align}
which can be proven straightforwardly.
Consequently, we find for the dc-conductivity
\begin{align}
\begin{aligned}
\sigma_{\alpha\beta}^\text{Bastin}&=\frac{e^2\pi\hbar}{V}\int\limits_{-\infty}^\infty dE\,f(E)\frac{d}{dE}\text{Tr}\left[\delta(E-\hat{H})\hat{v}_\alpha\delta(E-\hat{H})\hat{v}_\beta\right]+\sigma_\text{H},
\end{aligned}
\end{align}
where the Hall conductivity is obtained from the anti-symmetric term $C^\text{as}_{\alpha\beta}(E)$ in the Kubo-Bastin formula.
Using integration by parts for the energy integral involving $C^\text{s}_{\alpha\beta}(E)$ in the Kubo-Bastin formula, we readily obtain the equilibrium value of the dc-conductivity of Eq.~\eqref{equilibrium value dc-conductivity},
\begin{align}
\begin{aligned}
\sigma_{\alpha\beta}^\text{Bastin}
&=\sigma_{\alpha\beta}^\text{s}+\sigma_\text{H}\\
&=\sigma_{\alpha\beta},
\end{aligned}
\end{align}
which is equivalent to Eq.~\eqref{DC-conductivity} in the limit $\tau\to\infty$.
We furthermore can split the antisymmetric term $C_{\alpha\beta}^\text{as}$ into two terms yielding
\begin{align}
\begin{aligned}
C_{\alpha\beta}^\text{as}
&=-\frac{1}{2}\text{Tr}\left[\frac{d}{dE}\delta(E-\hat{H})\left(\hat{v}_\alpha\hat{x}_\beta-\hat{v}_\beta\hat{x}_\alpha\right)\right]+\text{Tr}\left[\delta(E-\hat{H})\left(\hat{v}_\alpha\left[\Re\hat{G}(E),\hat{x}_\beta\right]+h.c.\right)\right].
\end{aligned}
\end{align}

Hence, the Hall conductivity is decomposed into
\begin{align}
\sigma_\text{H}=\sigma_\text{H}^\text{I}+\sigma_\text{H}^\text{II},
\end{align}
with 
\begin{align}
&\begin{aligned}
\sigma_\text{H}^\text{I}&=\frac{e^2}{2V}\int\limits_{-\infty}^\infty dE\,f(E)\text{Tr}\left[\delta(E-\hat{H})\left(\hat{v}_\alpha\left[\Re\hat{G}(E),\hat{x}_\beta\right]+h.c.\right)\right],
\end{aligned}\\
&\sigma_\text{H}^\text{II}=\frac{e^2}{2V}\int\limits_{-\infty}^\infty dE\,f(E)\text{Tr}\left[\frac{d}{dE}\delta(E-\hat{H})\left(\hat{x}_\alpha\hat{v}_\beta-\hat{x}_\beta\hat{v}_\alpha\right)\right].
\end{align}
We note that $\sigma_\text{H}^\text{II}$ also appears in the St\v{r}eda-formula \cite{Streda_1982}. The above decomposition of the Hall conductivity can be used to prove Eq.~\eqref{sigma_H sc} (see Appendix \ref{app:C}), i.e., the semi-classical result for the Hall conductivity.

\section{\uppercase{Material Parameters of the effective Polaron model with vibrational disorder}}\label{app:F}

\begin{table}[h!]
	\centering
	\begin{tabular}{rcr}
		\hline\hline
		material parameter  && value  \\
		   &&  (meV) \\
		\hline
		onsite energy &&  0.0 \\
		$\varepsilon_{\text{AA}+\mathbf{b}}$ && 134.0\\
		$\varepsilon_{\text{AA}+2\mathbf{b}}$  &&  -10.7\\
		$\varepsilon_{\text{AC}}$ && 28.6\\
		$\varepsilon_{\text{AB}}$ && 4.1\\
		$V_0^{300\text{ K}}$  &&  52.1\\
		$V_{\text{AA}+\mathbf{b}}^{300\text{ K}}$ &&  20.9\\
		$V_{\text{AA}+2\mathbf{b}}^{300\text{ K}}$ &&  0.0\\
		$V_{\text{AC}}^{300\text{ K}}$ &&  8.7\\
		$V_{\text{AB}}^{300\text{ K}}$ && 0.9\\
		$V_0^{140\text{ K}}$  &&  48.2\\
		$V_{\text{AA}+\mathbf{b}}^{140\text{ K}}$ &&  15.1\\
		$V_{\text{AA}+2\mathbf{b}}^{140\text{ K}}$ &&  0.0 \\
		$V_{\text{AC}}^{140\text{ K}}$ &&  6.3\\
		$V_{\text{AB}}^{140\text{ K}}$ && 0.6\\
		\hline\hline
	\end{tabular}
	\caption{Summary of material parameters used in the effective polaron Hamiltonian in Eq.~\eqref{effective polaron Hamiltonian}. The values for the vibrational disorder are shown for the highest and for the lowest temperatures that have been investigated throughout this study. The molecular parameters for the electronic transfer integrals and the vibrational disorder are based on an \textit{ab initio} study of the electron-phonon coupling in rubrene. \cite{Ordejon2017} }
	\label{tab:material parameters}
\end{table}

\end{document}